\documentclass[preprintnumbers,amsmath,amssymb,nofootinbib,superscriptaddress,floatfix,onecolumn,reprint,groupedaddress,eqsecnum,amsfonts,aps]{revtex4}
\pdfoutput=1
\usepackage[normalem]{ulem}
\usepackage{graphicx}
\usepackage[colorlinks,citecolor=blue,urlcolor=blue,linkcolor=blue]{hyperref}
\usepackage{subfigure}
\usepackage{latexsym}
\usepackage{amssymb}
\usepackage{mathrsfs}
\usepackage{amsmath}
\usepackage{amsthm}
\usepackage{color}
\usepackage{dcolumn}
\usepackage{bm}

\definecolor{dyellow}{rgb}{1.,0.8,.0}
\definecolor{myblue}{rgb}{.1,.1,.7}
\definecolor{dcyan}{rgb}{.0,.6,.6}
\definecolor{dmagenta}{rgb}{0.6,0.0,0.6}
\definecolor{brown}{rgb}{0.6,0.2,0.}
\definecolor{darkblue}{rgb}{.0,.0,0.5}
\definecolor{darkred}{rgb}{0.75,0.0,0.0}
\definecolor{orange}{rgb}{1.,.6,.0}
\definecolor{dorange}{rgb}{0.8,.4,.0}
\definecolor{darkgreen}{rgb}{0.0,0.6,0.0}
\definecolor{purple}{rgb}{.4,.0,.4}
\definecolor{grey}{rgb}{0.5,0.5,0.5}

\begin{document}
\hyphenpenalty=1000
\title{Multipole analysis on gyroscopic precession in $f(R)$ gravity with irreducible Cartesian tensors}

\author{Bofeng Wu}\email{wubofeng@mail.neu.edu.cn}
\affiliation{Department of Physics, College of Sciences, Northeastern University, Shenyang 110819, China}

\author{Xin Zhang}
\email{zhangxin@mail.neu.edu.cn}
\affiliation{Department of Physics, College of Sciences, Northeastern University, Shenyang 110819, China}
\affiliation{Key Laboratory of Data Analytics and Optimization
for Smart Industry (Northeastern University), Ministry of Education, Shenyang 110819, China}

\begin{abstract}

In $f(R)$ gravity, the metric, presented in the form of the multipole expansion, for the external gravitational field of a spatially compact supported source up to $1/c^3$ order is provided, where $c$ is the velocity of light in vacuum. The metric consists of General Relativity-like part and $f(R)$ part, where the latter is the correction to the former in $f(R)$ gravity. At the leading pole order, the metric can reduce to that for a point-like or ball-like source.
For the gyroscope moving around the source without experiencing any torque, the multipole expansions of its spin's angular velocities of gravitoelectric-type precession, gravitomagnetic-type precession, $f(R)$ precession, and Thomas precession are all derived. The first two types of precession are collectively called General Relativity-like precession, and the $f(R)$ precession is the correction in $f(R)$ gravity. At the leading pole order, these expansions can recover the results for the gyroscope moving around a point-like or ball-like source. If the gyroscope has a nonzero four-acceleration, its spin's total angular velocity of precession up to $1/c^3$ order in $f(R)$ gravity is the same as that in General Relativity.
\end{abstract}
\maketitle
\section{Introduction}

So far, General Relativity (GR) has managed to survive many tests~\cite{Clifford2018}, and in particular, its predictions about gravitational waves are consistent with the recent observations by LIGO Scientific Collaboration and Virgo Collaboration~\cite{TheLIGOScientific:2016agk,TheLIGOScientific:2016agk1}, which shows that GR is a successful theory of gravity. Even so, there are still some observational data that cannot be well interpreted by GR at astrophysical and cosmological scales~\cite{cosmicacceleration}. In order to understand recent astronomical observations, the concepts of dark matter and dark energy have been introduced~\cite{Stabile:2010mz,Naf:2010zy}. Another approach to handling these challenges is to modify the Einstein's gravity theory~\cite{Sotiriou2010}. The metric $f(R)$ gravity~\cite{Gibbons1977,Starobinsky:1980te,Nojiri:2010wj,DNojiri:2017ncd} is a typical relativistic gravity theory, where the Einstein-Hilbert Lagrangian density of GR is replaced by a general function of Ricci scalar $R$.

The symmetric and trace-free (STF) formalism in terms of the irreducible Cartesian tensors, developed by Thorne~\cite{Thorne:1980ru}, Blanchet, and Damour~\cite{Blanchet:1985sp,Blanchet:1989ki}, is one of the important methods with respect to the multipole expansion, which can be used to describe the external gravitational field of the source localized in a finite region of space~\cite{Damour:1990gj}. In Ref.~\cite{Wu:2017huang}, the method of STF formalism is applied to the linearized $f(R)$ gravity, and its multipole expansion is presented explicitly in the far-field region,
so that the far-field metric outside a spatially compact supported source is obtained.

In this paper, for the gyroscope moving around a spatially compact supported source, we shall make a multipole analysis on its spin's angular velocity of precession in $f(R)$ gravity with the STF formalism. To this end, firstly, we should derive the metric in the whole region exterior to the source in $f(R)$ gravity, so that gyroscopic precession could be studied in a more general case. In addition, as the case in GR~\cite{MTW1973}, under the weak-field and slow-motion (WFSM) approximation, the metric for the external gravitational field of the source only needs to be expanded up to $1/c^3$ order, where $1/c$ is used as the WFSM parameter~\cite{Blanchet:2013haa}, so the linearized $f(R)$ gravity is sufficient to be used to analyze the gyroscopic precession~\cite{Dass:2019kon,Dass:2019hnb}.

In fact, it is the condition ``up to $1/c^3$ order" that greatly simplifies the derivation. By following the method in Ref.~\cite{Wu:2017huang}, the metric, presented in the form of the multipole expansion, for the external gravitational field of a spatially compact supported source up to $1/c^3$ order is derived under the de Donder condition in the present paper, and it consists of GR-like part and $f(R)$ part, where the former is exactly the result in GR when $f(R)$ gravity reduces to GR, and the latter is the correction to GR-like part in $f(R)$ gravity. When the leading pole moments are considered in the stationary spacetime, the GR-like part can recover the Lense-Thirring metric, and the $f(R)$ part provides the Yukawa-like correction in $f(R)$ gravity, so the metric can easily reduce to that for a point-like source in Refs.~\cite{Naf:2010zy,Dass:2019kon,Dass:2019hnb}. Further, if the leading pole moments are considered in the static spacetime, the metric can also reduce to that for a ball-like source in Ref.~\cite{Castel-Branco:2014exa}.

Following the conventional method in Ref.~\cite{MTW1973}, the calculation of the precessional angular velocity of gyroscopic spin in the stationary spacetime can be performed easily. However, the metric obtained above for the external gravitational field of the source is normally time-dependent, so the method in Ref.~\cite{MTW1973} should be extended. In the present paper, for the gyroscope moving around a spatially compact supported source without experiencing any torque, the multipole expansions of its spin's angular velocities of
gravitoelectric-type precession, gravitomagnetic-type precession, $f(R)$ precession, and Thomas precession, are all derived. The first two types of precession, associated with the mass-type and current-type source multipole moments of the GR-like part of the metric, are collectively called GR-like precession, which is
the result in GR when gyroscope moves along a geodesic. The $f(R)$ precession, associated with the source multipole moments of the $f(R)$ part of the metric, provides the correction in $f(R)$ gravity. The Thomas precession consists of the corresponding result in Special Relativity and the correction to this result brought about by the curved spacetime in $f(R)$ gravity.
It can be proved that if the gyroscope has a nonzero four-acceleration, its spin's total angular velocity of precession up to $1/c^3$ order in $f(R)$ gravity is the same as that in GR.

The four multipole expansions obtained above describe all the effects of the external gravitational field of the source up to $1/c^3$ order on gyroscopic precession, and in general, since the source multipole moments are time-dependent, the precessional angular velocities of gyroscopic spin are also time-dependent. When the effects at the leading pole order are considered in the stationary spacetime, the gravitoelectric-monopole effect, the gravitomagnetic-dipole effect, the $f(R)$-monopole effect, and the Thomas-monopole effect
are given, respectively. The first two effects, as those of the GR-like precession, can indeed recover classical geodesic effect and Lense-Thirring effect in GR, respectively. The $f(R)$-monopole effect provides the most main correction in $f(R)$ gravity, and it can reduce to that for the gyroscope moving around a point-like~\cite{Naf:2010zy,Dass:2019hnb} or a ball-like source~\cite{Castel-Branco:2014exa}.  The Thomas-monopole effect gives the most main correction to the result in Special Relativity. Further, by analogy, the effects at the next-leading and higher pole order can also be read out. In Refs.~\cite{Naf:2010zy,Castel-Branco:2014exa}, by comparing the effects of the gravitoelectric-type precession plus $f(R)$ precession at the leading pole order with the measurements of the experiment Gravity Probe B (GP-B), the constraints on the coefficient of the quadratic term in the Lagrangian density of $f(R)$ gravity are obtained. In this process, if the effects at the next-leading and higher pole order are considered further, one will acquire the influence of the scale and shape of the source (the Earth) on gyroscopic precession, so that a more accurate result may be obtained.

This paper is organized as follows. In Sec.~\ref{Sec:second}, the STF formalism and the metric $f(R)$ gravity are briefly reviewed. In Sec.~\ref{Sec:third}, the metric for the external gravitational field of a spatially compact supported source up to $1/c^3$ order is derived. In Sec.~\ref{Sec:fourth}, for the gyroscope moving around the source, the multipole expansions of its spin's angular velocities of precession in $f(R)$ gravity are obtained. In Sec.~\ref{Sec:fifth}, the conclusions and the related discussions are presented.

As in Ref.~\cite{Wu:2017huang}, the international system of units is used throughout this paper. When the notation is concerned, the Greek letters denote spacetime indices and range from 0 to 3, whereas the Latin letters denote space indices and range from 1 to 3. The repeated indices within a term represent that the sum should be taken over.
\section{Preliminary\label{Sec:second}}
\subsection{Relevant notations and formulas in the STF formalism~\label{Sec:STF}}
The knowledge of STF formalism is presented in detail in Ref.~\cite{Damour:1990gj}, and here, only the relevant notations and formulas are shown. In the linearized gravity theory, the coordinates $(x^{\mu})=(ct,x_{i})$ are regarded as the Minkowskian coordinates. The spherical coordinate system $(ct,r,\theta,\varphi)$ is defined by
\begin{equation}\label{equ2.1}
x_{1}=r\sin{\theta}\cos{\varphi},\ x_{2}=r\sin{\theta}\sin{\varphi},\ x_{3}=r\cos{\theta}.
\end{equation}
In the flat space, the radial vector is $\boldsymbol{x}=x_{i}\partial_{i}$, where $x_{i}$ are the components, and $\partial_{i}:=\partial/\partial x_{i}$ are the coordinate basis vectors. The unit radial vector is $\boldsymbol{n}=\boldsymbol{x}/r$, where $r=|\boldsymbol{x}|$ is the length of $\boldsymbol{x}$, and then, by defining $n_{i}=x_{i}/r$, there is $\boldsymbol{n}=n_{i}\partial_{i}$. Obviously, from Eq.~(\ref{equ2.1}),
\begin{equation}\label{equ2.2}
\partial_{r}:=\frac{\partial}{\partial r}=n_{i}\partial_{i}=\boldsymbol{n}.
\end{equation}

Given a Cartesian tensor with $l$ indices $B_{I_{l}}:=B_{i_{1}i_{2}\cdots i_{l}}$~\cite{Thorne:1980ru},
its STF part is
\begin{eqnarray}
\label{equ2.3}
\hat{B}_{I_{l}}:=B_{<I_{l}>}=B_{<i_{1}i_{2}\cdots i_{l}>}
:=\sum_{k=0}^{\left[\frac{l}{2}\right]}b_{k}\delta_{(i_{1}i_{2}}\cdots\delta_{i_{2k-1}i_{2k}}S_{i_{2k+1}\cdots i_{l})a_{1}a_{1}\cdots a_{k}a_{k}},
\end{eqnarray}
where
\begin{equation}
\label{equ2.4}b_{k}:=(-1)^{k}\frac{(2l-2k-1)!!}{(2l-1)!!}\frac{l!}{(2k)!!(l-2k)!},
\end{equation}
and
\begin{equation}\label{equ2.5}
S_{I_{l}}:=B_{(I_{l})}=B_{(i_{1}i_{2}\cdots i_{l})}:=\frac{1}{l!}\sum_{\sigma} B_{i_{\sigma(1)}i_{\sigma(2)}\cdots i_{\sigma(l)}}
\end{equation}
is its symmetric part with $\sigma$ running over all permutations of $(12\cdots l)$. The tensor products of $l$ radial and unit radial vectors are abbreviated by
\begin{eqnarray}
\label{equ2.6}X_{I_{l}}&=&X_{i_{1}i_{2}\cdots i_{l}}:= x_{i_{1}}x_{i_{2}}\cdots x_{i_{l}},\\
\label{equ2.7}N_{I_{l}}&=&N_{i_{1}i_{2}\cdots i_{l}}:= n_{i_{1}}n_{i_{2}}\cdots n_{i_{l}}
\end{eqnarray}
with
\begin{equation}\label{equ2.8}
X_{I_{l}}=r^l N_{I_{l}}.
\end{equation}

In addition, there are
\begin{eqnarray}
\label{equ2.9}\hat{N}_{I_{l}}&=&\sum_{k=0}^{\left[\frac{l}{2}\right]}b_{k}\delta_{(i_{1}i_{2}}\cdots\delta_{i_{2k-1}i_{2k}}
N_{i_{2k+1}\cdots i_{l})},\\
\label{equ2.10}\hat{\partial}_{I_{l}}&=&\sum_{k=0}^{\left[\frac{l}{2}\right]}b_{k}\delta_{(i_{1}i_{2}}\cdots\delta_{i_{2k-1}i_{2k}}
\partial_{i_{2k+1}\cdots i_{l})}\left(\nabla^2\right)^k,\\
\label{equ2.11}\hat{\partial}_{I_{l}}\left(\frac{F(r)}{r}\right)&=&\hat{N}_{I_{l}}\sum_{k=0}^{l}\frac{(l+k)!}{(-2)^{k}k!(l-k)!}
\frac{\partial_{r}^{l-k}F(r)}{r^{k+1}},
\end{eqnarray}
where $\nabla^2=\partial_{a}\partial_{a}$ is the Laplace operator in a flat space, $\partial_{I_{l}}=\partial_{i_{1}i_{2}\cdots i_{l}}:=\partial_{i_{1}}\partial_{i_{2}}\cdots\partial_{i_{l}}$,
and $\partial_r^{l-k}$ is the $(l-k)$-th derivative with respect to $r$.
\subsection{Metric $f(R)$ gravity~\label{Sec:f(R)gravity}}
Consider the spacetime with the metric $g_{\mu\nu}$ of signature $(-,+,+,+)$. The action of the metric $f(R)$ gravity~\cite{Wu:2017huang} is
\begin{equation}\label{equ2.12}
S=\frac{1}{2\kappa c}\int dx^4\sqrt{-g}f(R)+S_{M}(g^{\mu\nu},\psi),
\end{equation}
where $\kappa=8\pi G/c^{4}$ with $G$ as the gravitational constant, $g$ is the determinant of metric $g_{\mu\nu}$, and $S_{M}(g^{\mu\nu},\psi)$ is the matter action. The variation of the above action with respect to metric $g^{\mu\nu}$ yields the gravitational field equations
\begin{equation}\label{equ2.13}
H_{\mu\nu}=\kappa T_{\mu\nu}
\end{equation}
with
\begin{equation}
\label{equ2.14}H_{\mu\nu}:=-\frac{g_{\mu\nu}}{2}f+(R_{\mu\nu}+g_{\mu\nu}\square-\nabla_{\mu}\nabla_{\nu})f_{R}
\end{equation}
where $f_{R}:=\partial_{R}f:=\partial/\partial R$, and $T_{\mu\nu}$
is the energy momentum tensor. As in Ref.~\cite{Wu:2017huang}, $f(R)$ is assumed to have the polynomial form
\begin{equation}\label{equ2.15}
f(R)=R+a R^{2}+b R^{3}+\cdots,
\end{equation}
where the dimensions of the constants $a,b\cdots$ are $[R]^{-1},[R]^{-2}\cdots$, respectively.
\section{Metric for the external gravitational field of a spatially compact supported source up to $1/c^3$ order~\label{Sec:third}}
In Ref.~\cite{Wu:2017huang}, for a spatially compact supported source, the multipole analysis on linearized $f(R)$ gravity with the STF formalism is made
in a fictitious flat spacetime with $\eta^{\mu\nu}$ as the Minkowskian metric. Firstly, by defining the gravitational field amplitude $h^{\mu\nu}$ and the effective gravitational field amplitude $\tilde{h}^{\mu\nu}$ as
\begin{eqnarray}
\label{equ3.1}h^{\mu\nu}&:=&\sqrt{-g}g^{\mu\nu}-\eta^{\mu\nu},\\
\label{equ3.2}\tilde{h}^{\mu\nu}&:=&f_{R}\sqrt{-g}g^{\mu\nu}-\eta^{\mu\nu},
\end{eqnarray}
the field equations of $f(R)$ gravity are rewritten as
\begin{equation}\label{equ3.3}
\square_{\eta} \tilde{h}^{\mu\nu}=2\kappa\tau^{\mu\nu}_{f}
\end{equation}
under the de Donder condition $\partial_{\mu}\tilde{h}^{\mu\nu}=0$, where
$\square_{\eta}:=\eta^{\mu\nu}\partial_{\mu}\partial_{\nu}$, and
the source term
\begin{equation}\label{equ3.4}
\tau^{\mu\nu}_{f}:=|g|f_{R}^{2}T^{\mu\nu}+\frac{1}{2\kappa}\Lambda^{\mu\nu}_{f}
\end{equation}
is the stress-energy pseudotensor of the matter fields and the gravitational field. Here,
\begin{eqnarray}
\label{equ3.5}\Lambda^{\mu\nu}_{f}&:=&-\tilde{h}^{\alpha\beta}\partial_{\alpha}\partial_{\beta}\tilde{h}^{\mu\nu}
-(f_{R}-1)\tilde{g}^{\alpha\beta}\partial_{\alpha}\tilde{h}^{\mu\nu}\partial_{\beta}\ln{f_{R}}
-\frac{1}{2}(1+2f_{R})\tilde{g}^{\mu\nu}\tilde{g}^{\alpha\beta}\partial_{\alpha}\ln{f_{R}}\partial_{\beta}\ln{f_{R}}\notag\\
&&-(1-4f_{R})\tilde{g}^{\mu\alpha}\tilde{g}^{\nu\beta}\partial_{\alpha}\ln{f_{R}}\partial_{\beta}\ln{f_{R}}
-2(f_{R}-1)\tilde{g}^{\mu\nu}\tilde{g}^{\alpha\beta}\partial_{\alpha}\partial_{\beta}\ln{f_{R}}
+2(f_{R}-1)\tilde{g}^{\mu\alpha}\tilde{g}^{\nu\beta}\partial_{\alpha}\partial_{\beta}\ln{f_{R}}\notag\\
&&-2\tilde{g}_{\beta\tau}\tilde{g}^{\alpha(\mu}\partial_{\lambda}\tilde{h}^{\nu)\tau}\partial_{\alpha}\tilde{h}^{\beta\lambda}
-(f_{R}-1)\tilde{g}_{\rho\sigma}\tilde{g}^{\mu(\alpha}\tilde{g}^{\beta)\nu}\partial_{\alpha}\tilde{h}^{\rho\sigma}\partial_{\beta}\ln{f_{R}}
+\tilde{g}_{\alpha\beta}\tilde{g}^{\lambda\tau}\partial_{\lambda}\tilde{h}^{\mu\alpha}\partial_{\tau}\tilde{h}^{\nu\beta}\notag\\
&&+\partial_{\alpha}\tilde{h}^{\mu\beta}\partial_{\beta}\tilde{h}^{\nu\alpha}
-2(1-f_{R})\tilde{g}^{\alpha(\mu}\partial_{\alpha}\tilde{h}^{\nu)\beta}\partial_{\beta}\ln{f_{R}}
-\frac{1}{2}(1-f_{R})\tilde{g}_{\rho\sigma}\tilde{g}^{\mu\nu}\tilde{g}^{\alpha\beta}\partial_{\alpha}\tilde{h}^{\rho\sigma}\partial_{\beta}\ln{f_{R}}\notag\\
&&+\frac{1}{2}\tilde{g}_{\alpha\beta}\tilde{g}^{\mu\nu}\partial_{\lambda}\tilde{h}^{\alpha\tau}\partial_{\tau}\tilde{h}^{\beta\lambda}
+\frac{1}{8}(2\tilde{g}^{\mu\alpha}\tilde{g}^{\nu\beta}
-\tilde{g}^{\mu\nu}\tilde{g}^{\alpha\beta})(2\tilde{g}_{\lambda\tau}\tilde{g}_{\epsilon\pi}-\tilde{g}_{\epsilon\tau}\tilde{g}_{\lambda\pi})
\partial_{\alpha}\tilde{h}^{\lambda\pi}\partial_{\beta}\tilde{h}^{\tau\epsilon}\notag\\
&&+a\sqrt{-g}f_{R}\tilde{g}^{\mu\nu}R^{2}+b\sqrt{-g}f_{R}\tilde{g}^{\mu\nu}R^{3}+4agf_{R}^{2}R^{\mu\nu}R+6bgf_{R}^{2}R^{\mu\nu}R^{2}+\mbox{higher order terms}
\end{eqnarray}
with
\begin{eqnarray}
\label{equ3.6}\tilde{g}^{\mu\nu}&:=&f_{R}\sqrt{-g}g^{\mu\nu},\\
\label{equ3.7}\tilde{g}_{\mu\nu}&:=&\frac{1}{\sqrt{-g}f_{R}}g_{\mu\nu},
\end{eqnarray}
which satisfy
\begin{equation}
\label{equ3.8}\tilde{g}_{\mu\lambda}\tilde{g}^{\lambda\nu}=\delta^{\nu}_{\mu}.
\end{equation}
In the linearized $f(R)$ gravity, both $h^{\mu\nu}$ and $\tilde{h}^{\mu\nu}$ are the perturbations, namely,
\begin{eqnarray}
\label{equ3.9}|h^{\mu\nu}|&\ll &1,\\
\label{equ3.10}|\tilde{h}^{\mu\nu}|&\ll &1,
\end{eqnarray}
and the linearized relation between them is
\begin{equation}
\label{equ3.11}h^{\mu\nu}=\tilde{h}^{\mu\nu}-2aR^{(1)}\eta^{\mu\nu},
\end{equation}
where $a$ is the coefficient of the quadratic term in the Lagrangian density of $f(R)$ gravity, and $R^{(1)}$ is the linear part of Ricci scalar $R$. Eq.~(\ref{equ3.11}) shows that the gravitational field amplitude $h^{\mu\nu}$ consists of the tensor part, associated with $\tilde{h}^{\mu\nu}$, and the scalar part, associated with $R^{(1)}$, where $\tilde{h}^{\mu\nu}$ satisfies the following system of linear equations,
\begin{equation}\label{equ3.12}
\left\{\begin{array}{ll}
\displaystyle \square_{\eta}\tilde{h}^{\mu\nu}&=\displaystyle 2\kappa T^{\mu\nu},\smallskip\\
\displaystyle \partial_{\mu}\tilde{h}^{\mu\nu}&=\displaystyle 0,
\end{array}\right.
\end{equation}
and $R^{(1)}$ satisfies a massive Klein-Gordon equation with an external
source,
\begin{equation}
\label{equ3.13}\square_{\eta} R^{(1)}-m^{2}R^{(1)}=m^{2}\kappa T
\end{equation}
with $T:=\eta_{\mu\nu}T^{\mu\nu}$ and
\begin{equation}
\label{equ3.14}m^{2}:=\frac{1}{6a}.
\end{equation}

From Eq.~(\ref{equ3.1}), the pure information of the metric is carried by the gravitational field amplitude $h^{\mu\nu}$, so
its expression up to $1/c^3$ order interests us. Eq.~(\ref{equ3.11}) implies that
the expressions of $\tilde{h}^{\mu\nu}$ and $R^{(1)}$ up to $1/c^3$ order should be derived firstly. In Ref.~\cite{Wu:2017huang}, the multipole expansion of $\tilde{h}^{\mu\nu}$ in linearized $f(R)$ gravity is
\begin{equation}\label{equ3.15}
\left\{\begin{array}{ll}
\displaystyle\tilde{h}^{00}(t,\boldsymbol{x})&=\displaystyle -\frac{4G}{c^{2}}\sum_{l=0}^{\infty}\frac{(-1)^{l}}{l!}\partial_{I_{l}}\left( \frac{\hat{M}_{I_{l}}(u)}{r}\right),\smallskip\\
\displaystyle\tilde{h}^{0i}(t,\boldsymbol{x})&=\displaystyle \frac{4G}{c^{3}}\sum_{l=1}^{\infty}\frac{(-1)^{l}}{l!}\partial_{I_{l-1}}\left(\frac{\partial_{t}\hat{M}_{iI_{l-1}}(u)}{r}\right) +\frac{4G}{c^{3}}\sum_{l=1}^{\infty}\frac{(-1)^{l}l}{(l+1)!}\epsilon_{iab}\partial_{aI_{l-1}}\left( \frac{\hat{S}_{bI_{l-1}}(u)}{r}\right),\smallskip\\
\displaystyle\tilde{h}^{ij}(t,\boldsymbol{x})&=\displaystyle-\frac{4G}{c^{4}}\sum_{l=2}^{\infty}\frac{(-1)^{l}}{l!}\partial_{I_{l-2}}\left(\frac{\partial_{t}^{2}\hat{M}_{ijI_{l-2}}(u)}{r}\right) -\frac{8G}{c^{4}}\sum_{l=2}^{\infty}\frac{(-1)^{l}l}{(l+1)!}\partial_{aI_{l-2}}\left(\frac{\epsilon_{ab(i}\partial_{t}\hat{S}_{j)bI_{l-2}}(u)}{r}\right),
\end{array}\right.
\end{equation}
where $\epsilon_{ijk}$ is the totally antisymmetric Levi-Civita symbol with $\epsilon_{123}=1$,
\begin{equation}\label{equ3.16}
\left\{\begin{array}{ll}
\displaystyle\hat{M}_{I_{l}}(u)=&\displaystyle \frac{1}{c^{2}}\int d^{3}x'\left( \hat{X'}_{I_{l}}\left(\overline{T}^{00}_{l}(u,\boldsymbol{x}')+\overline{T}^{aa}_{l}(u,\boldsymbol{x}')\right)
-\frac{4(2l+1)}{c(l+1)(2l+3)}\hat{X'}_{aI_{l}}\partial_{t}\overline{T}^{0a}_{l+1}(u,\boldsymbol{x}') \right . \\
\displaystyle &\displaystyle\left .
+\frac{2(2l+1)}{c^2(l+1)(l+2)(2l+5)}\hat{X'}_{abI_{l}}\partial_{t}^{2}\overline{T}^{ab}_{l+2}(u,\boldsymbol{x}')\right ),\smallskip\\
\displaystyle\hat{S}_{I_{l}}(u)=&\displaystyle \frac{1}{c}\int d^{3}x'\left(\epsilon_{ab<i_{1}}\hat{X'}_{|a|i_{2}\cdots i_{l}>}\overline{T}^{0b}_{l}(u,\boldsymbol{x}') \right . \\
\displaystyle &\displaystyle\left . -\frac{2l+1}{c(l+2)(2l+3)}\epsilon_{ab<i_{1}}\hat{X'}_{|ac|i_{2}\cdots i_{l}>}\partial_{t}\overline{T}^{cb}_{l+1}(u,\boldsymbol{x}')\right),\ l\geq1
\end{array}\right.
\end{equation}
are the mass-type and current-type source multipole moments, respectively~\cite{Blanchet:2013haa},
$u=t-r/c$ is the retarded time, $\partial_t^{k}$ is the $k$-th derivative with respect to $t$, the symbol $<i_{1}|a|i_{2}\cdots i_{l}>$ and $<i_{1}|ac|i_{2}\cdots i_{l}>$
represent that $a$ and $c$ are not the STF indices, and
\begin{equation}
\label{equ3.17}\overline{T}^{\mu\nu}_{l}(u,\boldsymbol{x}'):=\frac{(2l+1)!!}{2^{l+1}l!}\int_{-1}^{1}(1-z^2)^{l}T^{\mu\nu}\left(u+\frac{zr'}{c},
\boldsymbol{x}'\right)dz
\end{equation}
with $r'=|\boldsymbol{x}'|$ as the length of $\boldsymbol{x}'$. By directly truncating the above multipole expansion of $\tilde{h}^{\mu\nu}$ under the WFSM approximation, its expression up to $1/c^3$ order can be readily obtained,
\begin{equation}\label{equ3.18}
\left\{\begin{array}{ll}
\displaystyle\tilde{h}^{00}(t,\boldsymbol{x})&=\displaystyle -\frac{4G}{c^{2}}\sum_{l=0}^{\infty}\frac{(-1)^{l}}{l!}\hat{M}_{I_{l}}(t)\partial_{I_{l}}\left( \frac{1}{r}\right),\smallskip\\
\displaystyle\tilde{h}^{0i}(t,\boldsymbol{x})&=\displaystyle \frac{4G}{c^{3}}\sum_{l=1}^{\infty}\frac{(-1)^{l}}{l!}\left(\partial_{t}\hat{M}_{iI_{l-1}}(t)\right)\partial_{I_{l-1}}\left(\frac{1}{r}\right) -\frac{4G}{c^{3}}\sum_{l=1}^{\infty}\frac{(-1)^{l}l}{(l+1)!}\epsilon_{iab}\hat{S}_{aI_{l-1}}(t)\partial_{bI_{l-1}}\left( \frac{1}{r}\right),\smallskip\\
\displaystyle\tilde{h}^{ij}(t,\boldsymbol{x})&=\displaystyle 0
\end{array}\right.
\end{equation}
with
\begin{equation}\label{equ3.19}
\left\{\begin{array}{ll}
\displaystyle\hat{M}_{I_{l}}(t)&=\displaystyle \int d^{3}x' \hat{X'}_{I_{l}}\frac{T^{00}(t,\boldsymbol{x}')}{c^2},\smallskip\\
\displaystyle\hat{S}_{I_{l}}(t)&=\displaystyle \int d^{3}x'\epsilon_{ab<i_{1}}\hat{X'}_{|a|i_{2}\cdots i_{l}>}\frac{T^{0b}(t,\boldsymbol{x}')}{c},\quad l\geq1,
\end{array}\right.
\end{equation}
where in this process~\cite{Damour:1990gj}, the conclusion
\begin{equation}
\label{equ3.20}T^{00}\sim O\big(c^2\big),\quad T^{0i}\sim O\big(c^1\big),\quad T^{ij}\sim O\big(c^0\big),
\end{equation}
the following series form of $\overline{T}^{\mu\nu}_{l}$,
\begin{equation}
\label{equ3.21}\overline{T}^{\mu\nu}_{l}(u,\boldsymbol{x}')=\sum_{k=0}^{\infty}\frac{(2l+1)!!}{(2k)!!(2l+2k+1)!!}
\frac{r'^{2k}}{c^{2k}}\frac{\partial^{2k}}{\partial u^{2k}}T^{\mu\nu}(u,\boldsymbol{x}'),
\end{equation}
and the conservation of the total mass of the source, namely,
\begin{equation}
\label{equ3.22}M:=\hat{M}_{I_{0}}(t)=\int d^{3}x' \frac{T^{00}(t,\boldsymbol{x}')}{c^2}
\end{equation}
have been used. Plugging Eq.~(\ref{equ3.18}) into Eq.~(\ref{equ3.12}) gives the system of equations satisfied by $\tilde{h}^{\mu0}$ under the WFSM approximation,
\begin{equation}\label{equ3.225}
\left\{\begin{array}{ll}
\displaystyle \square_{\eta}\tilde{h}^{\mu0}&=\displaystyle 2\kappa T^{\mu0},\smallskip\\
\displaystyle \partial_{\mu}\tilde{h}^{\mu0}&=\displaystyle 0,
\end{array}\right.
\end{equation}
which is analogous to Maxwell equations and Lorentz gauge condition, so $\tilde{h}^{00}$ and $\tilde{h}^{0i}$ should be related to some kind of gravitoelectric potential and gravitomagnetic vector potential~\cite{Ruggiero:2002hz}, respectively, and thus, $\tilde{h}^{00}$ and $\tilde{h}^{0i}$ could be called the gravitoelectric and gravitomagnetic components of the gravitational field amplitude, respectively.

Eqs.~(\ref{equ3.18}) and (\ref{equ3.19}) show that gravitoelectric component $\tilde{h}^{00}(t,\boldsymbol{x})$ is only associated with the mass-type source multipole moments, whereas gravitomagnetic components $\tilde{h}^{0i}(t,\boldsymbol{x})$ are associated with both mass-type and current-type source multipole moments. In the stationary case, there are
\begin{equation}
\label{equ3.23}T^{\mu\nu}(t',\boldsymbol{x}')=T^{\mu\nu}(\boldsymbol{x}'),
\end{equation}
and substituting them in Eqs.~(\ref{equ3.18}) and (\ref{equ3.19}) shows that $\tilde{h}^{0i}(\boldsymbol{x})$ are no longer associated with the mass-type source multipole moments,
which means that $\tilde{h}^{00}(\boldsymbol{x})$ and $\tilde{h}^{0i}(\boldsymbol{x})$ are decoupled in this case. From Eqs.~(\ref{equ2.15}) and (\ref{equ3.11}),
when $f(R)$ gravity reduces to GR, namely,
\begin{equation}
\label{equ3.24}f(R)=R,
\end{equation}
the above expression of $\tilde{h}^{\mu\nu}$ up to $1/c^3$ order is exactly the corresponding result of $h^{\mu\nu}$ in GR, also defined by Eq.~(\ref{equ3.1}), and therefore, the effective gravitational field amplitude $\tilde{h}^{\mu\nu}(t,\boldsymbol{x})$ is referred to as the GR-like part of $h^{\mu\nu}$~\cite{Wu:2017huang}.
Besides, when the leading pole moments are taken into account in the stationary spacetime, Eq.~(\ref{equ3.18}) becomes
\begin{equation}\label{equ3.25}
\left\{\begin{array}{ll}
\displaystyle\tilde{h}^{00}(\boldsymbol{x})&=\displaystyle -\frac{4GM}{c^{2}r},\smallskip\\
\displaystyle\tilde{h}^{0i}(\boldsymbol{x})&=\displaystyle \frac{2G\epsilon_{iab}x_{a}J_{b}}{c^{3}r^3},\smallskip\\
\displaystyle\tilde{h}^{ij}(\boldsymbol{x})&=\displaystyle 0
\end{array}\right.
\end{equation}
with
\begin{equation}\label{equ3.26}
J_{b}:=\hat{S}_{b}=\int d^{3}x'\epsilon_{bij}x'_{i}\frac{T^{0j}(\boldsymbol{x}')}{c}
\end{equation}
as the conserved angular momentum of the source, where if Eq.~(\ref{equ3.24}) holds, and the source is rotating around the $z$-axis, Eq.~(\ref{equ3.25}), as shown by following Eqs.~(\ref{equ3.41}) and (\ref{equ3.42}), can recover the Lense-Thirring metric in the isotropic coordinate.

The mulitpole expansion of $R^{(1)}$ is also provided in Ref.~\cite{Wu:2017huang}, and however, it is valid only in the far-field region. As mentioned before, gyroscopic precession is expected to be discussed in the whole region exterior to the source, so the expression of $R^{(1)}$ up to $1/c^3$ order can not be directly obtained like that of $\tilde{h}^{\mu\nu}$. In this section, we will rederive the multipole expansion of $R^{(1)}$ outside the source by imposing the condition ``up to $1/c^3$ order". Under the WFSM approximation, expansion of $R^{(1)}$ up to $1/c^3$ order means
\begin{equation}
\label{equ3.27}R^{(1)}\sim O\left(\frac{1}{c^2}\right),
\end{equation}
and then, Eq.~(\ref{equ3.13}) reduces to~\cite{Naf:2010zy}
\begin{equation}
\label{equ3.28}\nabla^{2}R^{(1)}-m^{2}R^{(1)}=\frac{8\pi Gm^{2}}{c^4} T,
\end{equation}
where from Eq.~(\ref{equ3.20}),
\begin{equation}
\label{equ3.29}T=-T^{00}\sim O\big(c^2\big).
\end{equation}
Eq.~(\ref{equ3.28}) has the following solution~\cite{Naf:2010zy}
\begin{equation}
\label{equ3.30}R^{(1)}(t,\boldsymbol{x})=\int\mathcal{G}(\boldsymbol{x};\boldsymbol{x}')\left(-\frac{8\pi Gm^{2}}{c^4} T(t,\boldsymbol{x}')\right)d^{3}x'
\end{equation}
with the Green's function
\begin{equation}
\label{equ3.31}\mathcal{G}(\boldsymbol{x};\boldsymbol{x}')=\frac{\text{e}^{-m|\boldsymbol{x}-\boldsymbol{x}'|}}{4\pi|\boldsymbol{x}-\boldsymbol{x}'|}.
\end{equation}
According to the result in Ref.~\cite{Wu:2017huang}, the above Green's function can be written as
\begin{equation}
\label{equ3.32}\mathcal{G}(\boldsymbol{x};\boldsymbol{x}')=\sum_{l=0}^{\infty}\frac{(2l+1)!!}{4\pi l!}mi_{l}(mr_<)k_{l}(mr_> )\hat{N}_{I_{l}}(\theta,\varphi)\hat{N}_{I_{l}}(\theta',\varphi'),
\end{equation}
where
\begin{equation}
\label{equ3.33}i_{l}(z):=z^l\left(\frac{d}{zdz}\right)^{l}\left(\frac{\sinh{z}}{z}\right),\qquad
k_{l}(z):=\frac{\text{e}^{-z}}{z}\sum_{k=0}^{l}\frac{(l+k)!}{k!(l-k)!}\frac{1}{(2z)^{k}}
\end{equation}
are the spherical modified Bessel functions of $l$-order~\cite{Arfken1985}, $(\theta,\varphi)$ and $(\theta',\varphi')$ are the angle coordinates of $\boldsymbol{x}$ and $\boldsymbol{x}'$, respectively, $r_{<}$ represents the lesser of $r=|\boldsymbol{x}|$ and $r'=|\boldsymbol{x}'|$, and $r_{>}$ the greater.
With the help of Eq.~(\ref{equ3.33}) and $$\text{e}^{-z}=(-1)^{l-k}\frac{d^{l-k}}{dz^{l-k}}\text{e}^{-z},$$ the multipole expansion of $R^{(1)}$ outside the source ($r=r_{>}$ and $r'=r_{<}$)
is given by inserting Eq.~(\ref{equ3.32}) into Eq.~(\ref{equ3.30}), namely,
\begin{eqnarray}
\label{equ3.34}R^{(1)}(t,\boldsymbol{x})\phantom{:}&=&-\frac{2Gm^2}{c^2}\sum_{l=0}^{\infty}\frac{(-1)^l}{l!}
\hat{Q}_{I_{l}}(t)\hat{N}_{I_{l}}(\theta,\varphi)\sum_{k=0}^{l}\frac{(l+k)!}{(-2)^{k}k!(l-k)!}\frac{1}{r^{k+1}}\frac{d^{l-k}}{dr^{l-k}}\text{e}^{-mr},\\
\label{equ3.35}\hat{Q}_{I_{l}}(t):&=&\frac{(2l+1)!!}{m^{2l}}\int r'^l\left(\frac{d}{r'dr'}\right)^{l}\left(\frac{\sinh{(mr')}}{mr'}\right)\hat{N}_{I_{l}}(\theta',\varphi')\frac{T(t,\boldsymbol{x}')}{c^2}d^{3}x',
\end{eqnarray}
where $\hat{Q}_{I_{l}}(t)$ are the $l$-pole moments. Then, by using Eq.~(\ref{equ2.11}) and $X'_{I_{l}}={r'}^l N_{I_l}(\theta',\varphi')$, the expression of $R^{(1)}$ up to $1/c^3$ order is
\begin{eqnarray}
\label{equ3.36}R^{(1)}(t,\boldsymbol{x})
&=&-\frac{2Gm^2}{c^2}\sum_{l=0}^{\infty}\frac{(-1)^l}{l!}\hat{Q}_{I_{l}}(t)\partial_{I_{l}}\left(\frac{\text{e}^{-mr}}{r}\right)
\end{eqnarray}
with
\begin{eqnarray}
\label{equ3.37}\hat{Q}_{I_{l}}(t)&=&\frac{(2l+1)!!}{m^{2l}}\int \hat{X'}_{I_{l}}\left(\frac{d}{r'dr'}\right)^{l}\left(\frac{\sinh{(mr')}}{mr'}\right)\frac{T(t,\boldsymbol{x}')}{c^2}d^{3}x'.
\end{eqnarray}

From the above process, it can be seen that the condition ``up to $1/c^3$ order" greatly simplifies the derivation. Firstly, it is due to this condition that Eq.~(\ref{equ3.13}) reduces to Eq.~(\ref{equ3.28}) under the WFSM approximation, and then, because the Green's function $\mathcal{G}(\boldsymbol{x};\boldsymbol{x}')$ of differential equation (\ref{equ3.28}) is the same as that in the stationary spacetime~\cite{Naf:2010zy}, the expression of $R^{(1)}$ up to $1/c^3$ order, presented in the form of the multipole expansion, is readily derived according to the result of $\mathcal{G}(\boldsymbol{x};\boldsymbol{x}')$ in Ref.~\cite{Wu:2017huang}. Obviously, except that $R^{(1)}(t,\boldsymbol{x})$, $\hat{Q}_{I_{l}}(t)$, and $T(t,\boldsymbol{x}')$ are time-dependent, the above expression of $R^{(1)}$ up to $1/c^3$ order is identical to that in the stationary spacetime~\cite{Wu:2017huang}. By use of Eqs.~(\ref{equ3.14}) and (\ref{equ3.36}),
the expression of the scalar part of $h^{\mu\nu}$, $2aR^{(1)}\eta^{\mu\nu}$, up to $1/c^3$ order can be directly obtained, and then, by inserting it and Eq.~(\ref{equ3.18}) into Eq.~(\ref{equ3.11}),
the expression of the gravitational field amplitude $h^{\mu\nu}$ up to $1/c^3$ order is
\begin{equation}\label{equ3.38}
\left\{\begin{array}{ll}
\displaystyle h^{00}(t,\boldsymbol{x})&=\displaystyle -\frac{4G}{c^{2}}\sum_{l=0}^{\infty}\frac{(-1)^{l}}{l!}\hat{M}_{I_{l}}(t)\partial_{I_{l}}\left( \frac{1}{r}\right)-\frac{2G}{3c^2}\sum_{l=0}^{\infty}\frac{(-1)^l}{l!}\hat{Q}_{I_{l}}(t)\hat\partial_{I_{l}}\left(\frac{\text{e}^{-mr}}{r}\right),\smallskip\\
\displaystyle h^{0i}(t,\boldsymbol{x})&=\displaystyle \frac{4G}{c^{3}}\sum_{l=1}^{\infty}\frac{(-1)^{l}}{l!}\left(\partial_{t}\hat{M}_{iI_{l-1}}(t)\right)\partial_{I_{l-1}}\left(\frac{1}{r}\right) -\frac{4G}{c^{3}}\sum_{l=1}^{\infty}\frac{(-1)^{l}l}{(l+1)!}\epsilon_{iab}\hat{S}_{aI_{l-1}}(t)\partial_{bI_{l-1}}\left( \frac{1}{r}\right),\smallskip\\
\displaystyle h^{ij}(t,\boldsymbol{x})&=\displaystyle \frac{2G}{3c^2}\delta^{ij}\sum_{l=0}^{\infty}\frac{(-1)^l}{l!}\hat{Q}_{I_{l}}(t)\hat\partial_{I_{l}}\left(\frac{\text{e}^{-mr}}{r}\right)
\end{array}\right.
\end{equation}
with $\delta^{ij}$ as the Kronecker symbol, which shows that the scalar part of $h^{\mu\nu}$ is the correction to the tensor part, namely, the GR-like part.
When the leading pole moments are considered in the stationary spacetime, from Eqs.~(\ref{equ3.25}), (\ref{equ3.36}), and (\ref{equ3.37}), Eq.~(\ref{equ3.38}) becomes
\begin{equation}\label{equ3.39}
\left\{\begin{array}{ll}
\displaystyle h^{00}(\boldsymbol{x})&=\displaystyle  -\frac{4GM}{c^{2}r}-\frac{2GM_{f}}{3c^2r}\text{e}^{-mr},\smallskip\\
\displaystyle h^{0i}(\boldsymbol{x})&=\displaystyle \frac{2G\epsilon_{iab}x_{a}J_{b}}{c^{3}r^3},\smallskip\\
\displaystyle h^{ij}(\boldsymbol{x})&=\displaystyle \frac{2GM_{f}}{3c^2r}\text{e}^{-mr}\delta^{ij},
\end{array}\right.
\end{equation}
where
\begin{eqnarray}
\label{equ3.40}M_{f}&:=&\hat{Q}_{I_{0}}=\int\left(\frac{\sinh{(mr')}}{mr'}\right)\frac{T(\boldsymbol{x}')}{c^2}d^{3}x'
\end{eqnarray}
is the stationary monopole moment of $R^{(1)}$, and in this case, the scalar part of $h^{\mu\nu}$ reduces to the Yukawa-like correction to the tensor part.
In the linearized $f(R)$ gravity, from Eq.~(\ref{equ3.9}), the trace of the gravitational field amplitude is $h=\eta_{\mu\nu}h^{\mu\nu}$, and then, with Eq.~(\ref{equ3.1}),
the metric for the gravitational field is given by
\begin{eqnarray}
\label{equ3.41}g_{\mu\nu}=\eta_{\mu\nu}-\overline{h}_{\mu\nu}
\end{eqnarray}
with
\begin{eqnarray}
\label{equ3.42}\overline{h}_{\mu\nu}:=h_{\mu\nu}-\frac{1}{2}\eta_{\mu\nu}h.
\end{eqnarray}
Thus, by plugging Eq.~(\ref{equ3.38}) into Eqs.~(\ref{equ3.41}) and (\ref{equ3.42}), the metric, presented in the form of the multipole expansion, for the external gravitational field of a spatially compact supported source up to $1/c^3$ order is
\begin{equation}\label{equ3.43}
\left\{\begin{array}{ll}
\displaystyle g_{00}(t,\boldsymbol{x})&=\displaystyle -1+\frac{2}{c^{2}}U(t,\boldsymbol{x})+\frac{1}{c^{2}}V(t,\boldsymbol{x}),\smallskip\\
\displaystyle g_{0i}(t,\boldsymbol{x})&=\displaystyle -\frac{4}{c^{3}}U^{i}(t,\boldsymbol{x}),\smallskip\\
\displaystyle g_{ij}(t,\boldsymbol{x})&=\displaystyle \delta_{ij}\left(1+\frac{2}{c^{2}}U(t,\boldsymbol{x})-\frac{1}{c^{2}}V(t,\boldsymbol{x})\right),
\end{array}\right.
\end{equation}
where the potentials $U(t,\boldsymbol{x})$, $U^{i}(t,\boldsymbol{x})$, and $V(t,\boldsymbol{x})$ are, respectively, defined as
\begin{equation}\label{equ3.44}
\left\{\begin{array}{ll}
\displaystyle U(t,\boldsymbol{x})&:=\displaystyle G\sum_{l=0}^{\infty}\frac{(-1)^{l}}{l!}\hat{M}_{I_{l}}(t)\partial_{I_{l}}\left( \frac{1}{r}\right),\smallskip\\
\displaystyle U^{i}(t,\boldsymbol{x})&:=\displaystyle -G\sum_{l=1}^{\infty}\frac{(-1)^{l}}{l!}\left(\partial_{t}\hat{M}_{iI_{l-1}}(t)\right)\partial_{I_{l-1}}\left(\frac{1}{r}\right) +G\sum_{l=1}^{\infty}\frac{(-1)^{l}l}{(l+1)!}\epsilon_{iab}\hat{S}_{aI_{l-1}}(t)\partial_{bI_{l-1}}\left( \frac{1}{r}\right),\smallskip\\
\displaystyle V(t,\boldsymbol{x})&:=\displaystyle -\frac{2G}{3}\sum_{l=0}^{\infty}\frac{(-1)^l}{l!}\hat{Q}_{I_{l}}(t)\partial_{I_{l}}\left(\frac{\text{e}^{-mr}}{r}\right).
\end{array}\right.
\end{equation}
Eqs.~(\ref{equ2.15}) and (\ref{equ3.14}) show that the potential $V(t,\boldsymbol{x})$ will vanish when $f(R)$ gravity reduces to GR, and thus, the metric in Eq.~(\ref{equ3.43}) recovers the result in GR. Therefore, in the above expression of the metric, the terms, not related to the potential $V(t,\boldsymbol{x})$, constitute the GR-like part, and the remaining terms, only related to the potential $V(t,\boldsymbol{x})$, constitute the $f(R)$ part, which is the correction to the GR-like part in $f(R)$ gravity. Obviously, the GR-like part of the metric contains the mass-type and current-type source multipole moments associated with the tensor part of the gravitational field amplitude, whereas
the $f(R)$ part contains the source multipole moments associated with the scalar part of the gravitational field amplitude. In the stationary spacetime, above three potentials at the leading pole order reduce to
\begin{equation}\label{equ3.45}
\left\{\begin{array}{lll}
\displaystyle U(t,\boldsymbol{x})&=\displaystyle U(\boldsymbol{x})&=\displaystyle \frac{GM}{r},\smallskip\\
\displaystyle U^{i}(t,\boldsymbol{x})&=\displaystyle U^{i}(\boldsymbol{x})&=\displaystyle -\frac{G\epsilon_{iab}x_{a}J_{b}}{2r^3},\smallskip\\
\displaystyle V(t,\boldsymbol{x})&=\displaystyle V(\boldsymbol{x})&=\displaystyle -\frac{2GM_{f}}{3r}\text{e}^{-mr},
\end{array}\right.
\end{equation}
respectively, and then, Eq.~(\ref{equ3.43}) yields the corresponding metric in this case, namely,
\begin{equation}\label{equ3.46}
\left\{\begin{array}{ll}
\displaystyle g_{00}(\boldsymbol{x})&\displaystyle=-1+\frac{2GM}{c^{2}r}-\frac{2GM_{f}}{3c^2r}\text{e}^{-mr},\smallskip\\
\displaystyle g_{0i}(\boldsymbol{x})&\displaystyle=\frac{2G\epsilon_{iab}x_{a}J_{b}}{c^3r^3},\smallskip\\
\displaystyle g_{ij}(\boldsymbol{x})&\displaystyle=\delta_{ij}\left(1+\frac{2GM}{c^{2}r}+\frac{2GM_{f}}{3c^2r}\text{e}^{-mr}\right).
\end{array}\right.
\end{equation}
It is easy to verify that for the source rotating around the $z$-axis, the above metric can recover that for a point-like source in Refs.~\cite{Naf:2010zy,Dass:2019kon,Dass:2019hnb}.
Further, if the leading pole moments are considered in the static spacetime, the above metric can also recover that for a ball-like source in Ref.~\cite{Castel-Branco:2014exa}.
\section{Multipole expansions of the precessional angular velocities of gyroscopic spin in $f(R)$ gravity with the STF formalism~\label{Sec:fourth}}
In Ref.~\cite{MTW1973}, the conventional method with respect to the derivation of the precessional angular velocity of gyroscopic spin in the stationary spacetime is presented, and however, the metric obtained in Sec.~\ref{Sec:third} for the gravitational field of the source is normally time-dependent, so the method in Ref.~\cite{MTW1973} should be extended. In this section, for the gyroscope moving in the external gravitational field of a spatially compact supported source without experiencing any torque, we will derive the multipole expansion of its spin's angular velocity of precession in $f(R)$ gravity with the STF formalism. Consider the spacetime with $g_{\mu\nu}$ in Eq.~(\ref{equ3.43}) as the metric
and let $x^{\mu}(\tau)$ be the world line of the gyroscope with $\tau$ as the proper time. Gyroscopic four-velocity and spin (i.e., the angular momentum vector), denoted by $u^{\alpha}$ and $S^{\alpha}$, respectively, are always orthogonal to each other~\cite{Naf:2010zy}, namely,
\begin{eqnarray}
\label{equ4.1}u^{\alpha}S_{\alpha}=0.
\end{eqnarray}
Because gyroscopic spin $S^{\alpha}$ obeys Fermi-Walker transport along its world line $x^{\mu}(\tau)$~\cite{MTW1973}, the following transport equation
\begin{eqnarray}
\label{equ4.2}u^{\alpha}\nabla_{\alpha}S^{\beta}=\frac{dS^{\beta}}{d\tau}+u^{\alpha}S^{\lambda}\Gamma^{\beta}_{\lambda\alpha}=\frac{1}{c^2}a^{\rho}S_{\rho}u^{\beta}
\end{eqnarray}
holds, where $\nabla_{\alpha}$ denotes the covariant derivative, $\Gamma^{\beta}_{\lambda\alpha}$ is the Christoffel symbol, and $a^{\rho}$ is gyroscopic four-acceleration. With Eqs.~(\ref{equ4.1}) and (\ref{equ4.2}), there is
\begin{eqnarray}
\label{equ4.3}\frac{d(S^{\beta}S_{\beta})}{d\tau}=u^{\alpha}\nabla_{\alpha}(S^{\beta}S_{\beta})=2S_{\beta}u^{\alpha}\nabla_{\alpha}S^{\beta}=0,
\end{eqnarray}
which implies that $S^{\beta}S_{\beta}$ remains fixed along $x^{\mu}(\tau)$.

Now, let's review the fundamental process of evaluating precessional angular velocity of gyroscopic spin~\cite{MTW1973}. Firstly, the coordinate frame
\begin{eqnarray}
\label{equ4.4}\boldsymbol{g}_{\rho}:=\frac{\partial}{\partial x^{\rho}}
\end{eqnarray}
should be orthonormalized, so that a local orthonormal tetrad $\boldsymbol{e}_{[\sigma]}$, at rest in the coordinate frame, can be defined,
\begin{eqnarray}
\label{equ4.5}\boldsymbol{e}_{[\sigma]}:=A^{\rho}_{\phantom{\rho}[\sigma]}\boldsymbol{g}_{\rho},
\end{eqnarray}
where the Greek indices within square brackets are used to label the vectors of the tetrad $\boldsymbol{e}_{[\sigma]}$ and the components of a tensor with respect to this tetrad.
Denote the orthonormal frame comoving with the gyroscope by $\boldsymbol{e}_{(\alpha)}$, and then, as the gyroscope moves in the gravitational field, the local Lorentz boost $\Lambda^{[\sigma]}_{\phantom{\sigma}(\alpha)}$ from the local orthonormal tetrad $\boldsymbol{e}_{[\sigma]}$ to its comoving frame $\boldsymbol{e}_{(\alpha)}$ can be determined with its four-velocity $u^{\alpha}$,
\begin{eqnarray}
\label{equ4.6}\boldsymbol{e}_{(\alpha)}:=\Lambda^{[\sigma]}_{\phantom{[\sigma]}(\alpha)}\boldsymbol{e}_{[\sigma]},
\end{eqnarray}
where the Greek indices within round brackets, similar to the case of the square bracket, are used to label the vectors of the frame $\boldsymbol{e}_{(\alpha)}$ and the components of a tensor with respect to this frame. In addition, the orthonormality of $\boldsymbol{e}_{[\sigma]}$ and $\boldsymbol{e}_{(\alpha)}$ implies the Greek indices within parentheses should be raised and lowered with the Minkowski metric. Combining the above two transformations gives
\begin{eqnarray}
\label{equ4.7}\boldsymbol{e}_{(\alpha)}=\lambda^{\rho}_{\phantom{\rho}(\alpha)}\boldsymbol{g}_{\rho}
\end{eqnarray}
with
\begin{eqnarray}
\label{equ4.8}\lambda^{\rho}_{\phantom{\rho}(\alpha)}:=\Lambda^{[\sigma]}_{\phantom{[\sigma]}(\alpha)}A^{\rho}_{\phantom{\rho}[\sigma]},
\end{eqnarray}
where since the gyroscope is at rest in its comoving orthonormal frame $\boldsymbol{e}_{(\alpha)}$, the following equations
\begin{eqnarray}
\label{equ4.9}\lambda^{\rho}_{\phantom{\rho}(0)}&=& \frac{u^{\rho}}{c},\\
\label{equ4.10}\eta_{\alpha\beta}&=&\lambda^{\rho}_{\phantom{\rho}(\alpha)}\lambda^{\sigma}_{\phantom{\sigma}(\beta)}g_{\rho\sigma}
\end{eqnarray}
hold. Thus, once both $\Lambda^{[\sigma]}_{\phantom{\sigma}(\alpha)}$ and $A^{\rho}_{\phantom{\rho}[\sigma]}$ are derived, the comoving frame of the gyroscope, $\boldsymbol{e}_{(\alpha)}$, can be determined by the coordinate frame $\boldsymbol{g}_{\rho}$. For gyroscopic spin, Eq.~(\ref{equ4.7}) gives
\begin{eqnarray}
\label{equ4.11}S^{\beta}=\lambda^{\beta}_{\phantom{\beta}(\alpha)}S^{(\alpha)},
\end{eqnarray}
and then, together with Eqs.~(\ref{equ4.1}),  (\ref{equ4.9}), (\ref{equ4.10}), and
\begin{eqnarray}
\label{equ4.12}u_{\beta}u^{\beta}=-c^2,
\end{eqnarray}
one can deduce
\begin{eqnarray}
\label{equ4.13}S^{(0)}=0,
\end{eqnarray}
which means that gyroscopic spin is a purely spatial vector in its comoving frame. Further, Eq.~(\ref{equ4.3}) implies that the length of gyroscopic spin $S^{(i)}$ remains fixed along its world line $x^{\mu}(\tau)$, namely,
\begin{eqnarray}
\label{equ4.14}\frac{d\left(S^{(i)}S_{(i)}\right)}{d\tau}=\frac{d\left(S^{(i)}S^{(j)}\delta_{ij}\right)}{d\tau}=0,
\end{eqnarray}
and hence, $S^{(i)}$ always precesses relative to the comoving frame, which yields
\begin{eqnarray}
\label{equ4.15}\frac{dS^{(i)}}{d\tau}=\epsilon^{ijk}\omega^{(j)}S^{(k)}
\end{eqnarray}
with $\omega^{(j)}$ as the angular velocity of $S^{(i)}$. The objective of the derivation in this section is to write down the expression of $\omega^{(j)}$ up to $1/c^3$ order under the WFSM approximation.

Firstly, we need to derive the expression of $dS^{(i)}/d\tau$ up to $1/c^3$ order. With Eqs.~(\ref{equ4.11}) and (\ref{equ4.13}), we have
\begin{equation}\label{equ4.16}
S^{(\alpha)}=\mu^{(\alpha)}_{\phantom{(\alpha)}\beta}S^{\beta}\Rightarrow
\left\{\begin{array}{ll}
\displaystyle S^{(0)}&=\mu^{(0)}_{\phantom{(0)}\beta}S^{\beta}=0,\smallskip\\
\displaystyle S^{(i)}&=\mu^{(i)}_{\phantom{(i)}\beta}S^{\beta},
\end{array}\right.
\end{equation}
where $\mu^{(\alpha)}_{\phantom{\alpha}\beta}$ is the inverse transformation of $\lambda^{\beta}_{\phantom{\beta}(\alpha)}$, and they satisfy
\begin{eqnarray}\label{equ4.17}
\left\{\begin{array}{ll}
\displaystyle \mu^{(\alpha)}_{\phantom{(\alpha)}\rho}\ \lambda^{\rho}_{\phantom{\beta}(\beta)}&=\delta^{\alpha}_{\phantom{\alpha}\beta},\smallskip\\
\displaystyle \lambda^{\alpha}_{\phantom{\alpha}(\rho)}\ \mu^{(\rho)}_{\phantom{(\rho)}\beta}&=\delta^{\alpha}_{\phantom{\alpha}\beta}.
\end{array}\right.
\end{eqnarray}
In Appendix A, after deducing the expressions of both transformations, $A^{\rho}_{\phantom{\rho}[\sigma]}$ and $\Lambda^{[\sigma]}_{\phantom{\sigma}(\alpha)}$, up to $1/c^3$ order,
that of the compound transformation $\mu^{(\alpha)}_{\phantom{\alpha}\beta}$ is also given by use of Eqs.~(\ref{equ4.8}) and (\ref{equ4.17}),
\begin{equation}\label{equ4.18}
\left\{\begin{array}{ll}
\displaystyle \mu^{(0)}_{\phantom{(0)}0}&\displaystyle=1-\frac{1}{c^{2}}U(t,\boldsymbol{x})-\frac{1}{2c^{2}}V(t,\boldsymbol{x})+\frac{v^{k}v^{k}}{2c^2},\smallskip\\
\displaystyle \mu^{(0)}_{\phantom{(0)}i}&\displaystyle=-\frac{v^{i}}{c}-\frac{v^{i}v^{k}v^{k}}{2c^3}-\frac{3v^{i}}{c^3}U(t,\boldsymbol{x})+\frac{v^{i}}{2c^3}V(t,\boldsymbol{x})+\frac{4}{c^3}U^{i}(t,\boldsymbol{x}),\smallskip\\
\displaystyle \mu^{(j)}_{\phantom{(j)}0}&\displaystyle=-\frac{v^{j}}{c}-\frac{v^{j}v^{k}v^{k}}{2c^3}-\frac{v^{j}}{c^3}U(t,\boldsymbol{x})+\frac{v^{j}}{2c^3}V(t,\boldsymbol{x}),\smallskip\\
\displaystyle \mu^{(j)}_{\phantom{(i)}i}&\displaystyle=\left(1+\frac{1}{c^{2}}U(t,\boldsymbol{x})-\frac{1}{2c^{2}}V(t,\boldsymbol{x})\right)\delta_{ji}+\frac{v^{j}v^{i}}{2c^2}.
\end{array}\right.
\end{equation}
Plug Eq.~(\ref{equ4.18}) into Eq.~(\ref{equ4.16}), and then, the expressions of $S^{0}$ and $S^{(i)}$ up to $1/c^3$ order are, respectively,
\begin{eqnarray}
\label{equ4.19}S^{0}&=&S^{i}\left(\frac{v^{i}}{c}+\frac{4v^{i}}{c^3}U(t,\boldsymbol{x})-\frac{4}{c^3}U^{i}(t,\boldsymbol{x})\right),\\
\label{equ4.20}S^{(i)}&=&S^{i}\left(1+\frac{1}{c^{2}}U(t,\boldsymbol{x})-\frac{1}{2c^{2}}V(t,\boldsymbol{x})\right)-\frac{v^{i}v^{j}S^{j}}{2c^2},
\end{eqnarray}
where Eq.~(\ref{equ4.19}) has been used in the derivation of Eq.~(\ref{equ4.20}). Thus, the derivative of $S^{(i)}$ with respect to gyroscopic proper time $\tau$ is
\begin{eqnarray}
\label{equ4.21}\frac{dS^{(i)}}{d\tau}=&&\frac{dS^{i}}{d\tau}\left(1+\frac{1}{c^{2}}U(t,\boldsymbol{x})-\frac{1}{2c^{2}}V(t,\boldsymbol{x})\right)
+\frac{S^{i}}{c^{2}}\frac{d}{d\tau}\left(U(t,\boldsymbol{x})-\frac{1}{2}V(t,\boldsymbol{x})\right) \notag\\
&&-\frac{v^{j}S^{j}}{2c^2}\frac{dv^{i}}{d\tau}-\frac{v^{i}S^{j}}{2c^2}\frac{dv^{j}}{d\tau}-\frac{v^{i}v^{j}}{2c^2}\frac{dS^{j}}{d\tau},
\end{eqnarray}
where $\partial_{t}(U(t,\boldsymbol{x})-V(t,\boldsymbol{x})/2)$ will appear in the second term with $\partial_{t}:=\partial/\partial t$, and in general, since the metric for the spacetime is time-dependent, it will not vanish. In Eq.~(\ref{equ4.2}), $dS^{i}/d\tau$ has been provided,
\begin{eqnarray}
\label{equ4.22}\frac{dS^{i}}{d\tau}=-u^{\alpha}S^{\lambda}\Gamma^{i}_{\lambda\alpha}+\frac{1}{c^2}a^{\rho}S^{\sigma}g_{\rho\sigma}u^{i},
\end{eqnarray}
and as shown in Appendix B, under the WFSM approximation, its expression up to $1/c^3$ order is
\begin{eqnarray}
\label{equ4.23}\frac{dS^{i}}{d\tau}=&&\frac{2}{c^2}v^{j}S^{j}\partial_{i}U(t,\boldsymbol{x})-\frac{1}{c^2}v^{j}S^{i}\partial_{j}U(t,\boldsymbol{x})-\frac{1}{c^2}v^{i}S^{j}\partial_{j}U(t,\boldsymbol{x})-\frac{2}{c^2}S^{j}\left(\partial_{i}U^{j}(t,\boldsymbol{x})-\partial_{j}U^{i}(t,\boldsymbol{x})\right)\notag\\
&&+\frac{1}{2c^2}v^{j}S^{i}\partial_{j}V(t,\boldsymbol{x})+\frac{1}{2c^2}v^{i}S^{j}\partial_{j}V(t,\boldsymbol{x})-\frac{1}{c^2}S^{i}\partial_{t}\left(U(t,\boldsymbol{x})-\frac{1}{2}V(t,\boldsymbol{x})\right)+\frac{1}{c^2}a^{j}v^{i}S^{j}.
\end{eqnarray}
In the following, we shall see that $\partial_{t}(U(t,\boldsymbol{x})-V(t,\boldsymbol{x})/2)$ in Eqs.~(\ref{equ4.21}) and (\ref{equ4.23}) will cancel out, so
in the final result of the precessional angular velocity of gyroscopic spin, it will not appear. In addition, from Eq.~(\ref{equB7}),
\begin{eqnarray}
\label{equ4.24}\frac{1}{c^2}\frac{dv^{i}}{d\tau}&=&\frac{1}{c^2}a^{i}+\frac{1}{c^2}\partial_{i}U(t,\boldsymbol{x})+\frac{1}{2c^2}\partial_{i}V(t,\boldsymbol{x}),
\end{eqnarray}
and then, by substituting it and Eq.~(\ref{equ4.23}) in Eq.~(\ref{equ4.21}), the expression of $dS^{(i)}/d\tau$ up to $1/c^3$ order is obtained. With Eq.~(\ref{equ4.20}), $dS^{(i)}/d\tau$ can be rewritten in the form of Eq.~(\ref{equ4.15}), and then, the angular velocity of $S^{(i)}$ is given, namely,
\begin{eqnarray}
\label{equ4.25}\omega^{(j)}=\omega^{(j)}_{GR}+\omega^{(j)}_{f(R)}+\omega^{(j)}_{T}
\end{eqnarray}
with
\begin{eqnarray}
\label{equ4.26}\omega^{(j)}_{GR}&=&\frac{3}{2c^2}\epsilon^{jpq}v^{p}\partial_{q}U(t,\boldsymbol{x})+\frac{2}{c^2}\epsilon^{jpq}\partial_{p}U^{q}(t,\boldsymbol{x}),\\
\label{equ4.27}\omega^{(j)}_{f(R)}&=&-\frac{1}{4c^2}\epsilon^{jpq}v^{p}\partial_{q}V(t,\boldsymbol{x}),\\
\label{equ4.28}\omega^{(j)}_{T}&=&-\frac{1}{2c^2}\epsilon^{jpq}v^{p}a^{q}.
\end{eqnarray}
Eq.~(\ref{equ4.15}) describes the precession of gyroscopic spin $S^{(i)}$ relative to its comoving orthonormal frame $\boldsymbol{e}_{(\alpha)}$ in complete generality under the WFSM approximation, and $\omega^{(j)}_{GR}$, $\omega^{(j)}_{f(R)}$, and $\omega^{(j)}_{T}$ in Eq.~(\ref{equ4.25}) describe three types of precession, respectively. As mentioned previously, in terms of the metric, the potential $V(t,\boldsymbol{x})$ is associated with the $f(R)$ part, and it will vanish when $f(R)$ gravity reduces to GR, which then results in $\omega^{(j)}_{f(R)}=0$. Furthermore, if gyroscope has a vanishing four-acceleration, there is $\omega^{(j)}_{T}=0$. Thus, $\omega^{(j)}=\omega^{(j)}_{GR}$ means that $\omega^{(j)}_{GR}$ is the precessional angular velocity of gyroscopic spin in GR when gyroscope moves along a geodesic. The precession resulted from $\omega^{(j)}_{GR}$ is referred to as GR-like precession.
As to $\omega^{(j)}_{f(R)}$, it is related to the potential $V(t,\boldsymbol{x})$, so it is the corrected angular velocity of precession of gyroscopic spin in $f(R)$ gravity.  The precession resulted from $\omega^{(j)}_{f(R)}$ is referred to as $f(R)$ precession. Obviously, $\omega^{(j)}_{T}$ represents the Thomas precession, which plays an important role in the fine structure of atomic spectra~\cite{MTW1973}. By inserting Eqs.~(\ref{equ4.24}) and (\ref{equA4}) into Eq.~(\ref{equ4.28}), we get
\begin{eqnarray}
\label{equ4.29}\omega^{(j)}_{T}&=&-\frac{1}{2c^2}\epsilon^{jpq}v^{p}\frac{dv^{q}}{dt}+\frac{1}{2c^2}\epsilon^{jpq}v^{p}\partial_{q}U(t,\boldsymbol{x})+\frac{1}{4c^2}\epsilon^{jpq}v^{p}\partial_{q}V(t,\boldsymbol{x}),
\end{eqnarray}
where the first term is exactly the corresponding result in Special Relativity, and the last two terms, associated with the potentials $U(t,\boldsymbol{x})$ and $V(t,\boldsymbol{x})$, respectively, come from the connection coefficients in gyroscopic four-acceleration. In addition, although Eq.~(\ref{equ4.28}) implies that $\omega^{(j)}_{T}$ in $f(R)$ gravity has the same form as that in GR, they are actually completely different because the potential $V(t,\boldsymbol{x})$ is nonvanishing in $f(R)$ gravity. Next, we shall prove that if gyroscope has a nonzero four-acceleration,
$\omega^{(j)}$ actually has nothing to do with the potential $V(t,\boldsymbol{x})$.
With Eqs.~(\ref{equ4.25})---(\ref{equ4.27}) and (\ref{equ4.29}), $\omega^{(j)}$ can be rewritten as
\begin{eqnarray}
\label{equ4.30}\omega^{(j)}&=&-\frac{1}{2c^2}\epsilon^{jpq}v^{p}\frac{dv^{q}}{dt}+\frac{2}{c^2}\epsilon^{jpq}v^{p}\partial_{q}U(t,\boldsymbol{x})+\frac{2}{c^2}\epsilon^{jpq}\partial_{p}U^{q}(t,\boldsymbol{x}),
\end{eqnarray}
which shows that the potential $V(t,\boldsymbol{x})$ does not appear in the expression of $\omega^{(j)}$, and namely, the $f(R)$ part of the metric has nothing to do with $\omega^{(j)}$, so $\omega^{(j)}$ is also total angular velocity of precession of gyroscopic spin up to $1/c^3$ order in GR.

Let's focus on $\omega^{(j)}_{GR}$. Define
\begin{eqnarray}
\label{equ4.31}\omega^{(j)}_{GRE}&:=&\frac{3}{2c^2}\epsilon^{jpq}v^{p}\partial_{q}U(t,\boldsymbol{x}),\\
\label{equ4.32}\omega^{(j)}_{GRM}&:=&\frac{2}{c^2}\epsilon^{jpq}\partial_{p}U^{q}(t,\boldsymbol{x}),
\end{eqnarray}
and then,
\begin{eqnarray}
\label{equ4.33}\omega^{(j)}_{GR}=\omega^{(j)}_{GRE}+\omega^{(j)}_{GRM}.
\end{eqnarray}
Besides, with Eqs.~(\ref{equ3.18}) and (\ref{equ3.44}), we have
\begin{equation}\label{equ4.34}
\left\{\begin{array}{ll}
\displaystyle U(t,\boldsymbol{x})&=\displaystyle -\frac{c^2}{4}\tilde{h}^{00}(t,\boldsymbol{x}),\smallskip\\
\displaystyle U^{p}(t,\boldsymbol{x})&=\displaystyle -\frac{c^3}{4}\tilde{h}^{0p}(t,\boldsymbol{x}).\smallskip
\end{array}\right.
\end{equation}
Remember that $\tilde{h}^{00}$ and $\tilde{h}^{0i}$ are called the gravitoelectric and gravitomagnetic components of the gravitational field amplitude, respectively, so
two types of precession of gyroscopic spin resulted from $\omega^{(j)}_{GRE}$ and $\omega^{(j)}_{GRM}$ could be referred to as the gravitoelectric-type precession and gravitomagnetic-type precession, respectively. By inserting Eq.~(\ref{equ3.44}) into Eqs.~(\ref{equ4.31}), (\ref{equ4.32}), (\ref{equ4.27}), and (\ref{equ4.29}), the expressions of $\omega^{(j)}_{GRE}$, $\omega^{(j)}_{GRM}$, $\omega^{(j)}_{f(R)}$, and $\omega^{(j)}_{T}$ up to $1/c^3$ order, presented in the form of the multipole expansions, are all derived,
\begin{eqnarray}
\label{equ4.35}\omega^{(j)}_{GRE}&=&\frac{3G}{2c^2}\sum_{l=0}^{\infty}\frac{(-1)^{l}}{l!}\hat{M}_{I_{l}}(t)\epsilon^{jpq}v^{p}\partial_{qI_{l}}\left( \frac{1}{r}\right),\\
\label{equ4.36}\omega^{(j)}_{GRM}&=&\frac{2G}{c^2}\sum_{l=1}^{\infty}\frac{(-1)^{l}}{l!}\epsilon^{jpq}\left(\partial_{t}\hat{M}_{pI_{l-1}}(t)\right)\partial_{qI_{l-1}}\left(\frac{1}{r}\right)\notag\\ &&-\frac{2G}{c^2}\sum_{l=1}^{\infty}\frac{(-1)^{l}l}{(l+1)!}\hat{S}_{I_{l}}(t)\partial_{jI_{l}}\left( \frac{1}{r}\right),\\
\label{equ4.37}\omega^{(j)}_{f(R)}&=&\frac{G}{6c^2}\sum_{l=0}^{\infty}\frac{(-1)^l}{l!}\hat{Q}_{I_{l}}(t)\epsilon^{jpq}v^{p}\partial_{qI_{l}}\left(\frac{\text{e}^{-mr}}{r}\right),\\
\label{equ4.38}\omega^{(j)}_{T}&=&-\frac{1}{2c^2}\epsilon^{jpq}v^{p}\frac{dv^{q}}{dt}+\frac{G}{2c^2}\sum_{l=0}^{\infty}\frac{(-1)^{l}}{l!}\hat{M}_{I_{l}}(t)\epsilon^{jpq}v^{p}\partial_{qI_{l}}\left( \frac{1}{r}\right)\notag\\
&&-\frac{G}{6c^2}\sum_{l=0}^{\infty}\frac{(-1)^l}{l!}\hat{Q}_{I_{l}}(t)\epsilon^{jpq}v^{p}\partial_{qI_{l}}\left(\frac{\text{e}^{-mr}}{r}\right),
\end{eqnarray}
where in the derivation of the second term in Eq.~(\ref{equ4.36}), $\nabla^2(1/r)=0\ (r\neq0)$ has been used. Clearly, in general, since the source multipole moments are time-dependent,
above precessional angular velocities of gyroscopic spin are also time-dependent, and it is the dependence on time that results in that  $\omega^{(j)}_{GRM}$ is associated with the mass-type source multipole moments of the GR-like part of the metric. The above four multipole expansions describe all the effects of the external gravitational field of the source up to $1/c^3$ order on the gyroscopic precession. When it comes to the gyroscopic experiment, e.g., GP-B, the results in the stationary spacetime are significant, and in the following, we will discuss them. According to Eqs.~(\ref{equ3.19}) and (\ref{equ3.35}), the dependence of the source multipole moments on time disappears in the stationary spacetime, so there are
\begin{eqnarray}
\label{equ4.39}\omega^{(j)}_{GRE}&=&\frac{3G}{2c^2}\sum_{l=0}^{\infty}\frac{(-1)^{l}}{l!}\hat{M}_{I_{l}}\epsilon^{jpq}v^{p}\partial_{qI_{l}}\left( \frac{1}{r}\right),\\
\label{equ4.40}\omega^{(j)}_{GRM}&=&-\frac{2G}{c^2}\sum_{l=1}^{\infty}\frac{(-1)^{l}l}{(l+1)!}\hat{S}_{I_{l}}\partial_{jI_{l}}\left( \frac{1}{r}\right),\\
\label{equ4.41}\omega^{(j)}_{f(R)}&=&\frac{G}{6c^2}\sum_{l=0}^{\infty}\frac{(-1)^l}{l!}\hat{Q}_{I_{l}}\epsilon^{jpq}v^{p}\partial_{qI_{l}}\left(\frac{\text{e}^{-mr}}{r}\right),\\
\label{equ4.42}\omega^{(j)}_{T}&=&-\frac{1}{2c^2}\epsilon^{jpq}v^{p}\frac{dv^{q}}{dt}+\frac{G}{2c^2}\sum_{l=0}^{\infty}\frac{(-1)^{l}}{l!}\hat{M}_{I_{l}}\epsilon^{jpq}v^{p}\partial_{qI_{l}}\left( \frac{1}{r}\right)\notag\\
&&-\frac{G}{6c^2}\sum_{l=0}^{\infty}\frac{(-1)^l}{l!}\hat{Q}_{I_{l}}\epsilon^{jpq}v^{p}\partial_{qI_{l}}\left(\frac{\text{e}^{-mr}}{r}\right),
\end{eqnarray}
from which, we know that $\omega^{(j)}_{GRM}$ is no longer related to the mass-type source multipole moments of the GR-like part of the metric. Firstly, the effects at
the leading pole order need to be considered, and by truncating the above expansions, we get
\begin{eqnarray}
\label{equ4.43}\omega^{(j)}_{GRE}&=&-\frac{3GM}{2c^2r^3}\epsilon^{jpq}v^{p}x^{q},\\
\label{equ4.44}\omega^{(j)}_{GRM}&=&\frac{GJ_{i}}{c^2r^5}(3x^{i}x^{j}-\delta_{ij}r^2),\\
\label{equ4.45}\omega^{(j)}_{f(R)}&=&-\frac{GM_{f}(1+mr)\text{e}^{-mr}}{6c^2r^3}\epsilon^{jpq}v^{p}x^{q},\\
\label{equ4.46}\omega^{(j)}_{T}&=&-\frac{1}{2c^2}\epsilon^{jpq}v^{p}\frac{dv^{q}}{dt}-\frac{GM}{2c^2r^3}\epsilon^{jpq}v^{p}x^{q}\notag\\
&&+\frac{GM_{f}(1+mr)\text{e}^{-mr}}{6c^2r^3}\epsilon^{jpq}v^{p}x^{q}
\end{eqnarray}
with $M,J_{i}$, and $M_{f}$ defined in Eqs.~(\ref{equ3.22}), (\ref{equ3.26}), and (\ref{equ3.40}), where above $\omega^{(j)}_{GRE}$, $\omega^{(j)}_{GRM}$, $\omega^{(j)}_{f(R)}$, and $\omega^{(j)}_{T}$ represent the gravitoelectric-monopole effect, gravitomagnetic-dipole effect, $f(R)$-monopole effect, and Thomas-monopole effect, respectively. Obviously, the gravitoelectric-monopole effect is the geodesic effect, and when the source is rotating around the $z$-axis, the gravitomagnetic-dipole effect is the Lense-Thirring effect, where as those of the GR-like precession at the leading pole order, they can indeed recover the classical results in GR. The $f(R)$-monopole effect provides the most main correction in $f(R)$ gravity, and it can reduce to that for the gyroscope moving around a point-like~\cite{Naf:2010zy,Dass:2019hnb} or a ball-like source~\cite{Castel-Branco:2014exa}. As mentioned earlier, the first term in Eq.~(\ref{equ4.46}) is the Thomas precession in Special Relativity, and thus, the above Thomas-monopole effect, represented by the last two terms in Eq.~(\ref{equ4.46}), gives the most main correction to this result brought about by the curved spacetime in $f(R)$ gravity. Further, by analogy, the effects at the next-leading and higher pole order can also be read out.

{In the experiment GP-B, gyroscopes are moving around the Earth along the polar orbit~\cite{Everitt:2011hp}, and since the gyroscopes are in geodesic motion, one only needs to consider their gravitoelectric-type precession, gravitomagnetic-type precession, and $f(R)$ precession. In such a case, if only the effects
at the leading pole order are considered, Eqs.~(\ref{equ4.43})---(\ref{equ4.45}) show that the directions of $\omega^{(j)}_{GRE}+\omega^{(j)}_{f(R)}$ and $\omega^{(j)}_{GRM}$ are orthogonal to each other, so that these two types of effects can be handled separately. In addition, Eqs.~(\ref{equ4.43})---(\ref{equ4.45}) need to be averaged over a period~\cite{Naf:2010zy,Dass:2019hnb}, respectively, before compared with the experimental data. From Eqs.~(\ref{equ3.14}) and (\ref{equ4.45}), the angular velocity of $f(R)$ precession at the leading pole order contains the coefficient of the quadratic term in the Lagrangian density of $f(R)$ gravity, namely, $a$, so the constraint on $a$ can be obtained by matching the theoretical results of the gravitoelectric-monopole effect plus $f(R)$-monopole effect with the corresponding data from GP-B. In Ref.~\cite{Naf:2010zy}, for the gyroscope moving around a point-like source, the obtained constraint on $a$ is $|a|\lesssim5\times 10^{11}\text{m}^2$.}
In order to obtain a more accurate result, the influence of the scale and shape of the source (the Earth) on gyroscopic precession also needs to be considered. In Ref.~\cite{Castel-Branco:2014exa}, by using the result for a ball-like source at the leading pole order, the constraint on $a$ from the experiment GP-B does not yield any new result. In Ref.~\cite{Dass:2019hnb}, the influence of the Earth's oblateness on gyroscopic precession is considered, but the effective constraint on $a$ is not obtained.
{The results in the present paper provide an approach to effectively acquiring the influence of the scale and shape of a spatially compact supported source on gyroscopic precession. In principle, by comparing the results of gravitoelectric-type precession plus $f(R)$ precession at the next-leading and higher pole order with the data from GP-B or the future gyroscopic experiment, a more tight constraint on $a$ could be obtained. However, in this process, how to separate the effects of the gravitoelectric-type precession plus $f(R)$ precession from those of gravitomagnetic-type precession is still an open question, and moreover, taking the average of Eqs.~(\ref{equ4.39})---(\ref{equ4.41}) over a period is strenuous. Therefore, handling these issues is beyond the scope of this paper, and we will discuss them in future work. Anyway, what is confirmed is that the multipole expansions of the precessional angular velocities of gyroscopic spin presented in this paper are of great significance in application.}

\section{Conclusions and discussions \label{Sec:fifth}}

In this paper, for the gyroscope moving around a spatially compact supported source without experiencing any torque, the multipole analysis on its spin's angular velocity of precession in $f(R)$ gravity has been made under the WFSM approximation. The first problem that we confront is how to obtain the metric outside the source. Although the metric outside the source is explicitly presented in Ref.~\cite{Wu:2017huang} by applying the STF formalism, developed by Thorne~\cite{Thorne:1980ru}, Blanchet, and Damour~\cite{Blanchet:1985sp,Blanchet:1989ki}, to the linearized $f(R)$ gravity, it is only applicable in the far-field region. In order that gyroscopic precession could be studied in a more general case, the metric in the whole region exterior to the source needs to be derived. In addition, it needs to be emphasized that as the case in GR~\cite{MTW1973}, under the WFSM approximation, the metric for the gravitational field of the source only needs to be expanded up to $1/c^3$ order, so the linearized $f(R)$ gravity is sufficient to analyze the gyroscopic precession~\cite{Dass:2019kon,Dass:2019hnb}.

The condition ``up to $1/c^3$ order" does greatly simplifies the derivation. In the present paper, by following the STF formalism in Ref.~\cite{Wu:2017huang}, we derive the metric, presented in the form of the multipole expansion, for the external gravitational field of a spatially compact supported source up to $1/c^3$ order under the de Donder condition, and it consists of GR-like part and $f(R)$ part, where the former is exactly the result in GR when $f(R)$ gravity reduces to GR, and the latter is the correction to GR-like part in $f(R)$ gravity. In the stationary spacetime, the GR-like part of the metric at the leading pole order can recover the Lense-Thirring metric, and
the $f(R)$ part of the metric at the leading pole order yields the Yukawa-like correction in $f(R)$ gravity, so that the metric obtained can reduce to that for a point-like source in Refs.~\cite{Naf:2010zy,Dass:2019kon,Dass:2019hnb}. Further, in the static spacetime, the metric at the leading pole order can also reduce to that for a ball-like source in Ref.~\cite{Castel-Branco:2014exa}.

Since the metric obtained above for the external gravitational field of the source is normally time-dependent, the method in Ref.~\cite{MTW1973}, used for the calculation of
the precessional angular velocity of gyroscopic spin in the stationary spacetime, has been extended in this paper. Thus, after a detailed derivation, for the gyroscope moving around the source without experiencing any torque, we give all the multipole expansions of its spin's angular velocities of gravitoelectric-type precession, gravitomagnetic-type precession, $f(R)$ precession, and Thomas precession. The first two types of precession, as the results in GR when gyroscope moves along a geodesic, are collectively called GR-like precession, and they are associated with the mass-type and current-type source multipole moments of the GR-like part of the metric. The $f(R)$ precession provides the correction in $f(R)$ gravity, and it is associated with the source multipole moments of the $f(R)$ part of the metric. The Thomas precession consists of the corresponding result in Special Relativity and the correction to this result brought about by the curved spacetime in $f(R)$ gravity. We also give the proof that if the gyroscope has a nonzero four-acceleration, its spin's total angular velocity of precession in $f(R)$ gravity is the same as that in GR.

All the effects of the external gravitational field of the source up to $1/c^3$ order on gyroscopic precession are described by the four multipole expansions obtained above, and in general, since the source multipole moments are time-dependent, above precessional angular velocities of gyroscopic spin are also time-dependent. One might be more interested in the results in the stationary spacetime, when it comes to the gyroscopic experiment, and these results are also written down in this paper. In the stationary spacetime, the effects at the leading pole order consist of the gravitoelectric-monopole effect, the gravitomagnetic-dipole effect, the $f(R)$-monopole effect, and the Thomas-monopole effect. The first two effects are  those of GR-like precession at the leading pole order, and they can indeed recover classical geodesic effect and Lense-Thirring effect in GR, respectively. The $f(R)$-monopole effect provides the most main correction in $f(R)$ gravity, and it can reduce to that for the gyroscope moving around a point-like~\cite{Naf:2010zy,Dass:2019hnb} or a ball-like source~\cite{Castel-Branco:2014exa}. The Thomas-monopole effect gives the most main correction to the result in Special Relativity. Further, the effects at the next-leading and higher pole order can also be read out by analogy.

As far as we know, such expressions in $f(R)$ gravity, especially for the multipole expansions of the precessional angular velocities of gyroscopic spin, have not been given before. Although the results obtained in this paper, expanded only up to $1/c^3$ order, are derived in the theoretical framework of the linearized $f(R)$ gravity, they are sufficient to be used to explain the data from the experiment GP-B since the Earth's gravitational field is weak, and the gyroscope moves slowly. Further, if one wants to acquire the expression of the precessional angular velocity of gyroscopic spin up to $1/c^4$ order, from Eq.~(\ref{equ3.15}), we know that the quadratic terms of the effective gravitational field amplitude, like $(\tilde{h}^{00})^2$, also need to be taken into consideration. The metric $f(R)$ gravity is one of the simplest modified gravity theories. Although some models of $f(R)$ gravity can explain the inflation in early universe successfully~\cite{Nojiri:2010wj}, in order to better explain more observed phenomena, many generalized modified gravity theories, like $f(R,\mathcal{G})$ gravity~\cite{Shamir:2017ndy,Odintsov:2018nch} and $f(X,Y,Z)$ gravity~\cite{Stabile:2010mz}, are attracting considerable attention, where $\mathcal{G}$ is the Gauss-Bonnet invariant, $X:=R$ is the Ricci scalar, $Y:=R_{\mu\nu}R^{\mu\nu}$ is the quadratic contraction of two Ricci tensors, and $Z:=R_{\mu\nu\sigma\rho}R^{\mu\nu\sigma\rho}$ is the quadratic contraction of two Riemann tensors. The result for $f(R)$ gravity in this paper might be generalized to these theories, so that they can have a wide range of applications.
\begin{acknowledgments}
This work was supported by China Postdoctoral Science Foundation (Grant No. 2021M690569),
the National Natural Science Foundation of China (Grants Nos. 11975072, 11835009, 11875102, and 11690021), the Liaoning Revitalization Talents Program (Grant No. XLYC1905011), and the National Program for Support of Top-Notch Young Professionals (Grant No. W02070050).
\end{acknowledgments}

\appendix\label{appendix}
\section{DERIVATION OF EQ.~(\ref{equ4.18})}\label{appendix:1}
Eqs.~(\ref{equ4.8}) and (\ref{equ4.17}) show that in order to obtain the expression of the transformation $\mu^{(\alpha)}_{\phantom{(\alpha)}\beta}$ up to $1/c^3$ order,
those of both transformations, $A^{\rho}_{\phantom{\rho}[\sigma]}$ and $\Lambda^{[\sigma]}_{\phantom{\sigma}(\alpha)}$, need to be derived firstly. The metric for the external gravitational field of the source up to $1/c^3$ order has been presented in Eq.~(\ref{equ3.43}), and then, by applying the Gram-Schmidt orthogonalization to the coordinate frame $\boldsymbol{g}_{\rho}$ under the WFSM approximation, the local tetrad $\boldsymbol{e}_{[\sigma]}$ in Eq.~(\ref{equ4.5}) is
\begin{equation}\label{equA1}
\left\{\begin{array}{l}
\displaystyle \boldsymbol{e}_{[0]}=\left(1+\frac{1}{c^{2}}U(t,\boldsymbol{x})+\frac{1}{2c^{2}}V(t,\boldsymbol{x})\right)\boldsymbol{g}_{0},\smallskip\\
\displaystyle \boldsymbol{e}_{[i]}=-\frac{4}{c^3}U^{i}(t,\boldsymbol{x})\boldsymbol{g}_{0}+\left(1-\frac{1}{c^{2}}U(t,\boldsymbol{x})+\frac{1}{2c^{2}}V(t,\boldsymbol{x})\right)\boldsymbol{g}_{i},
\end{array}\right.
\end{equation}
from which, the expression of the transformation $A^{\rho}_{\phantom{\sigma}[\sigma]}$ up to $1/c^3$ order is obtained,
\begin{equation}\label{equA2}
\left\{\begin{array}{l}
\displaystyle A^{0}_{\phantom{0}[0]}=1+\frac{1}{c^{2}}U(t,\boldsymbol{x})+\frac{1}{2c^{2}}V(t,\boldsymbol{x}),\smallskip\\
\displaystyle A^{0}_{\phantom{0}[i]}=-\frac{4}{c^3}U^{i}(t,\boldsymbol{x}),\smallskip\\
\displaystyle A^{j}_{\phantom{j}[0]}=0,\smallskip\\
\displaystyle A^{j}_{\phantom{i}[i]}=\left(1-\frac{1}{c^{2}}U(t,\boldsymbol{x})+\frac{1}{2c^{2}}V(t,\boldsymbol{x})\right)\delta_{ji}.
\end{array}\right.
\end{equation}
Next, we will make use of gyroscopic four-velocity $u^{\alpha}$ to derive the Lorentz boost $\Lambda^{[\sigma]}_{\phantom{\sigma}(\alpha)}$ from the tetrad $\boldsymbol{e}_{[\sigma]}$ to the comoving frame $\boldsymbol{e}_{(\alpha)}$. Gyroscopic four-velocity $u^{\alpha}$ can be rewritten as
\begin{equation}\label{equA3}
\left\{\begin{array}{l}
\displaystyle u^{0}=c\frac{dt}{d\tau},\smallskip\\
\displaystyle u^{i}=v^{i}\frac{dt}{d\tau},
\end{array}\right.
\end{equation}
where $t=x^{0}/c$ is the coordinate time, and $v^{i}:=dx^{i}/dt$ is gyroscopic velocity measured using the coordinate frame. Then, due to Eq.~(\ref{equ4.12}),
the expression of $dt/d\tau$ up to $1/c^3$ order is
\begin{eqnarray}
\label{equA4}\frac{dt}{d\tau}=1+\frac{1}{c^{2}}U(t,\boldsymbol{x})+\frac{1}{2c^{2}}V(t,\boldsymbol{x})+\frac{1}{2c^{2}}v^{k}v^{k}.
\end{eqnarray}
Similarly, in the local tetrad $\boldsymbol{e}_{[\sigma]}$, gyroscopic four-velocity $u^{[\alpha]}$ can also be rewritten as
\begin{equation}\label{equA5}
\left\{\begin{array}{l}
\displaystyle u^{[0]}=c\frac{dt_{0}}{d\tau},\smallskip\\
\displaystyle u^{[i]}=v^{[i]}\frac{dt_{0}}{d\tau},
\end{array}\right.
\end{equation}
where $t_{0}$ is the proper time of the observer at rest in the tetrad $\boldsymbol{e}_{[\sigma]}$, and $v^{[i]}$ is gyroscopic velocity in this tetrad.
Eq.~(\ref{equ4.5}) yields
\begin{eqnarray*}
u^{[\sigma]}=\big(A^{-1}\big)^{[\sigma]}_{\phantom{[\sigma]}\rho}u^{\rho},
\end{eqnarray*}
and then, together with Eq.~(\ref{equA2})---(\ref{equA5}), we get
\begin{eqnarray}
\label{equA6}\frac{dt_{0}}{d\tau}&=&1+\frac{v^{k}v^{k}}{2c^2},\\
\label{equA7}v^{[i]}&=&\left(1+\frac{2}{c^{2}}U(t,\boldsymbol{x})\right)v^{i}.
\end{eqnarray}
As mentioned earlier, $v^{[i]}$ is gyroscopic velocity in the tetrad $\boldsymbol{e}_{[\sigma]}$, and then, with Eqs.~(\ref{equA6}) and (\ref{equA7}), the expression of the Lorentz boost $\Lambda^{[\sigma]}_{\phantom{\sigma}(\alpha)}$ up to $1/c^3$ order is
\begin{equation}\label{equA8}
\left\{\begin{array}{l}
\displaystyle \Lambda^{[0]}_{\phantom{[0]}(0)}=1+\frac{v^{k}v^{k}}{2c^2},\smallskip\\
\displaystyle \Lambda^{[0]}_{\phantom{[0]}(i)}=\frac{v^{i}}{c}+\frac{v^{i}v^{k}v^{k}}{2c^3}+\frac{2v^{i}}{c^3}U(t,\boldsymbol{x}),\smallskip\\
\displaystyle \Lambda^{[j]}_{\phantom{[j]}(0)}=\frac{v^{j}}{c}+\frac{v^{j}v^{k}v^{k}}{2c^3}+\frac{2v^{j}}{c^3}U(t,\boldsymbol{x}),\smallskip\\
\displaystyle \Lambda^{[j]}_{\phantom{[i]}(i)}=\delta_{ji}+\frac{v^{j}v^{i}}{2c^2}.
\end{array}\right.
\end{equation}
Plug Eqs.~(\ref{equA2}) and (\ref{equA8}) into Eq.~(\ref{equ4.8}), and then, from Eq.~(\ref{equ4.17}), the expression of the compound transformation $\mu^{(\alpha)}_{\phantom{(\alpha)}\beta}$ up to $1/c^3$ order is derived, namely, Eq.~(\ref{equ4.18}).
\section{DERIVATION OF EQ.~(\ref{equ4.23})}\label{appendix:2}
Firstly, from Eq.~(\ref{equ4.22}), the coefficients of the Levi-Civita connection, $\Gamma^{\beta}_{\lambda\alpha}$, need to be derived, and under the weak-field approximation, their definitions are
\begin{eqnarray}
\label{equB1}\Gamma^{\beta}_{\lambda\alpha}=\frac{1}{2}\bigg(\partial_{\alpha}\Big(-\overline{h}^{\beta}_{\phantom{\beta}\lambda}\Big)+\partial_{\lambda}\Big(-\overline{h}^{\beta}_{\phantom{\beta}\alpha}\Big)-\partial^{\beta}\Big(-\overline{h}_{\lambda\alpha}\Big)\bigg),
\end{eqnarray}
where $\partial^{\mu}:=\eta^{\mu\rho}\partial_{\rho}$. By using Eqs.~(\ref{equ3.41}) and (\ref{equ3.43}), we have
\begin{equation}\label{equB2}
\left\{\begin{array}{ll}
\displaystyle -\overline{h}_{00}(t,\boldsymbol{x})&=\displaystyle \frac{2}{c^{2}}U(t,\boldsymbol{x})+\frac{1}{c^{2}}V(t,\boldsymbol{x}),\smallskip\\
\displaystyle -\overline{h}_{0i}(t,\boldsymbol{x})&=\displaystyle -\frac{4}{c^{3}}U^{i}(t,\boldsymbol{x}),\smallskip\\
\displaystyle -\overline{h}_{ij}(t,\boldsymbol{x})&=\displaystyle \delta_{ij}\left(\frac{2}{c^{2}}U(t,\boldsymbol{x})-\frac{1}{c^{2}}V(t,\boldsymbol{x})\right),
\end{array}\right.
\end{equation}
and then, inserting them into Eq.~(\ref{equB1}) gives the expressions of the connection coefficients up to $1/c^3$ order,
\begin{equation}\label{equB3}
\left\{\begin{array}{ll}
\displaystyle \Gamma^{0}_{00}=&\displaystyle -\frac{1}{c^3}\partial_{t}U(t,\boldsymbol{x})-\frac{1}{2c^3}\partial_{t}V(t,\boldsymbol{x}),\smallskip\\
\displaystyle \Gamma^{0}_{0j}=&\displaystyle -\frac{1}{c^2}\partial_{j}U(t,\boldsymbol{x})-\frac{1}{2c^2}\partial_{j}V(t,\boldsymbol{x}),\smallskip\\
\displaystyle \Gamma^{0}_{jk}=&\displaystyle \frac{2}{c^3}\left(\partial_{j}U^{k}(t,\boldsymbol{x})+\partial_{k}U^{j}(t,\boldsymbol{x})\right)+\frac{\delta_{jk}}{c^3}\left(\partial_{t}U(t,\boldsymbol{x})-\frac{1}{2}\partial_{t}V(t,\boldsymbol{x})\right),\smallskip\\
\displaystyle \Gamma^{i}_{00}=&\displaystyle -\frac{1}{c^2}\partial_{i}U(t,\boldsymbol{x})-\frac{1}{2c^2}\partial_{i}V(t,\boldsymbol{x}),\smallskip\\
\displaystyle \Gamma^{i}_{0j}=&\displaystyle \frac{2}{c^3}\left(\partial_{i}U^{j}(t,\boldsymbol{x})-\partial_{j}U^{i}(t,\boldsymbol{x})\right)+\frac{\delta_{ij}}{c^3}\left(\partial_{t}U(t,\boldsymbol{x})-\frac{1}{2}\partial_{t}V(t,\boldsymbol{x})\right),\smallskip\\
\displaystyle \Gamma^{i}_{jk}=&\displaystyle \frac{1}{c^2}\big(\delta_{ij}\partial_{k}U(t,\boldsymbol{x})+\delta_{ik}\partial_{j}U(t,\boldsymbol{x})-\delta_{jk}\partial_{i}U(t,\boldsymbol{x})\big)\\
&\displaystyle -\frac{1}{2c^2}\big(\delta_{ij}\partial_{k}V(t,\boldsymbol{x})+\delta_{ik}\partial_{j}V(t,\boldsymbol{x})-\delta_{jk}\partial_{i}V(t,\boldsymbol{x})\big).
\end{array}\right.
\end{equation}
Thus, together with Eqs.~(\ref{equ4.19}), (\ref{equA3}), (\ref{equA4}), and (\ref{equB3}), the expression of the first term in Eq.~(\ref{equ4.22}) up to $1/c^3$ order is
\begin{eqnarray}
\label{equB4}-u^{\alpha}S^{\lambda}\Gamma^{i}_{\lambda\alpha}=&&\frac{2}{c^2}v^{j}S^{j}\partial_{i}U(t,\boldsymbol{x})-\frac{1}{c^2}v^{j}S^{i}\partial_{j}U(t,\boldsymbol{x})-\frac{1}{c^2}v^{i}S^{j}\partial_{j}U(t,\boldsymbol{x})-\frac{2}{c^2}S^{j}\left(\partial_{i}U^{j}(t,\boldsymbol{x})-\partial_{j}U^{i}(t,\boldsymbol{x})\right)\notag\\
&&+\frac{1}{2c^2}v^{j}S^{i}\partial_{j}V(t,\boldsymbol{x})+\frac{1}{2c^2}v^{i}S^{j}\partial_{j}V(t,\boldsymbol{x})-\frac{1}{c^2}S^{i}\left(\partial_{t}U(t,\boldsymbol{x})-\frac{1}{2}\partial_{t}V(t,\boldsymbol{x})\right).
\end{eqnarray}
Gyroscopic four-acceleration is
\begin{eqnarray}
\label{equB5}a^{\beta}=u^{\lambda}\nabla_{\lambda}u^{\beta}=\frac{du^{\beta}}{d\tau}+u^{\rho}u^{\sigma}\Gamma^{\beta}_{\rho\sigma},
\end{eqnarray}
and then, with Eqs.~(\ref{equA3}), (\ref{equA4}), and (\ref{equB3}), there are
\begin{eqnarray}
\label{equB6}\frac{1}{c^2}a^{0}&=&\frac{1}{2c^3}\frac{d(v^{k}v^{k})}{d\tau}-\frac{2}{c^3}v^{j}\partial_{j}U(t,\boldsymbol{x})-\frac{1}{c^3}v^{j}\partial_{j}V(t,\boldsymbol{x}),\\
\label{equB7}\frac{1}{c^2}a^{i}&=&\frac{1}{c^2}\frac{dv^{i}}{d\tau}-\frac{1}{c^2}\partial_{i}U(t,\boldsymbol{x})-\frac{1}{2c^2}\partial_{i}V(t,\boldsymbol{x}).
\end{eqnarray}
By plugging Eqs.~(\ref{equ3.43}), (\ref{equ4.19}), (\ref{equB6}), and (\ref{equB7}) into the second term in Eq.~(\ref{equ4.22}), its expression up to $1/c^3$ order is
\begin{eqnarray}
\label{equB8}\frac{1}{c^2}a^{\rho}S^{\sigma}g_{\rho\sigma}u^{i}=\frac{1}{c^2}a^{j}v^{i}S^{j}.
\end{eqnarray}
Thus, the expression of $dS^{i}/d\tau$ up to $1/c^3$ order can be obtained from Eqs.~(\ref{equB4}) and (\ref{equB8}), namely, Eq.~(\ref{equ4.23}).

\end{document}